%% file: paper.tex
\documentclass[a4paper,11pt]{article}

\usepackage[dvipsnames]{xcolor}
\usepackage{jheppub}
\usepackage{graphicx}
\usepackage{amsmath,amssymb,amsthm,amsfonts}
\usepackage{slashed}
\usepackage{braket}
\usepackage[export]{adjustbox}
\usepackage{enumitem}
\usepackage{placeins}
\usepackage[normalem]{ulem}
\usepackage{multirow}
\usepackage{hyperref}
\usepackage{mathrsfs}  
\usepackage{dsfont}

\usepackage{verbatim}
\usepackage[T1]{fontenc}
\usepackage[utf8]{inputenc}
\usepackage{array}
\usepackage{wrapfig}
\usepackage{tabu}
\usepackage{longtable}
\usepackage{float}
\usepackage{caption}
\usepackage{subcaption}
\usepackage{tikz,tikz-3dplot}
\usepackage[compat=1.0.0]{tikz-feynman}
\usetikzlibrary{calc,arrows.meta,shapes.geometric}
\usepackage{soul}
\usepackage[usestackEOL]{stackengine}
\usepackage{cancel}

\allowdisplaybreaks

\definecolor{mycolor}{rgb}{0, 0.73, 0.0}
\hypersetup{
    colorlinks=true,
    linkcolor=mycolor,
    citecolor=mycolor,
    filecolor=mycolor,      
    urlcolor=mycolor
    }

\def\beq{\begin{equation}}
\def\eeq{\end{equation}}
\def\bea{\begin{eqnarray}}
\def\eea{\end{eqnarray}}

\newcommand\greencheckmark[1][]{%
  \tikz[scale=0.4,#1]{\fill(0,.35) -- (.25,0) -- (1,.7) -- (.25,.15) -- cycle;}%
}
\newcommand\crossmark[1][]{%
  \tikz[scale=0.4,#1]{
    \fill(0,0)--(0.1,0) .. controls (0.5,0.4) .. (1,0.7)--(0.9,0.7) ..  controls (0.5,0.5) ..(0,0.1) --cycle;
    \fill(1,0.1)--(0.9,0.1) .. controls (0.5,0.3) .. (0,0.7)--(0.1,0.7) .. controls (0.5,0.4) ..(1,0.2) --cycle;
  }%
}

\preprint{SLAC-PUB-17742}

\title{Single-soft emissions for amplitudes with two colored particles at three loops}

\author[a]{Franz Herzog,}
\emailAdd{fherzog@ed.ac.uk}
\author[b]{Yao Ma,}
\emailAdd{yaomay@phys.ethz.ch}
\author[c]{Bernhard Mistlberger,}
\emailAdd{bernhard.mistlberger@gmail.com}
\author[c]{Adi Suresh}
\emailAdd{adisur@stanford.edu}

\affiliation[a]{Higgs Centre for Theoretical Physics, School of Physics and Astronomy,\\
The University of Edinburgh, Edinburgh EH9 3FD, Scotland, U.K.}
\affiliation[b]{Institute for Theoretical Physics, ETH Zürich, 8093 Zürich, Switzerland}
\affiliation[c]{SLAC National Accelerator Laboratory, Stanford University, Stanford, CA 94039, USA}

\abstract{
We compute the three-loop correction to the universal single-soft emission current for the case of scattering amplitudes with two additional color-charged partons. We present results valid for QCD and $\mathcal{N}=4$ super-symmetric Yang-Mills theory. To achieve our results we develop a new integrand expansion technique for scattering amplitudes in the presence of soft emissions. Furthermore, we obtain contributions from single final-state parton matrix elements to the Higgs boson and Drell-Yan production cross section at next-to-next-to-next-to-next-to leading order (N$^4$LO) in perturbative QCD in the threshold limit.
}

\begin{document}
\maketitle

\input{Chapters/Introduction.tex}
\input{Chapters/SetUp.tex}

\input{Chapters/Computation.tex}

\input{Chapters/Regions}
\input{Chapters/IRSubtraction.tex}
\input{Chapters/N4SYM.tex}
\input{Chapters/SoftXS.tex}

\input{Chapters/Conclusion.tex}
\acknowledgments
We thank Lance Dixon, Einan Gardi and Stephen Jones for useful discussions.
FH and YM are supported by the UKRI FLF grant ``Forest Formulas for the LHC'' (Mr/S03479x/1) and the STFC Consolidated Grant ``Particle Physics at the Higgs Centre''. BM and AS are supported by the United States Department of Energy, Contract DE-AC02-76SF00515.
YM would like to thank the Galileo Galilei Institute for Theoretical Physics for the hospitality and the INFN for partial support, during the completion of this work.
\appendix

\addcontentsline{toc}{section}{References}
\bibliographystyle{jhep}
\bibliography{refs}

\end{document}

%% file: Chapters/Introduction.tex
\section{Introduction}

A remarkable property of gauge theory scattering amplitudes is that they factorize in infrared limits. Infrared limits are generally characterized by soft and/or collinear momentum configurations, and typically lead to singularities or poles in the amplitude. In turn these singularities are responsible for the infrared divergences encountered in both loop and phase space integrals, which typically appear in intermediate stages of the computation of physical quantities.   

The infrared limits are of great interest from both a practical as well as theoretical perspective. For one, they are an important ingredient for building infrared subtraction schemes for higher-order QCD cross section calculations \cite{Ridder2005, Somogyi:2006da, Somogyi:2006db, Catani:2007vq, Czakon:2010td, Boughezal:2011jf, Gehrmann_De_Ridder_2012, Gaunt:2015pea, Caola:2017dug, Anastasiou:2018rib, Herzog:2018ily, Magnea:2018hab, Cieri:2018oms, Cieri_2019}. They are also responsible for potentially large logarithmic corrections in a variety of observables,  and as such enter as crucial ingredients for resummations \cite{Collins:1989gx,Sterman:1995fz,Bauer:2000ew,Bauer:2001yt}. Finally, infrared limits are of fundamental interest in the study of the mathematical properties of scattering amplitudes as they constrain their analytic structure. In this context, infrared limits have played an important role in the analytic bootstrap program \cite{Bern:1993qk,Almelid:2017qju,Caron-Huot:2019vjl}.

In this work, we will focus on the limit of scattering amplitudes involving three colored partons in which a single external gluon momentum becomes soft. 
It is well known that the all-order $n$-point amplitude factorizes in this limit into  an $n-1$-point amplitude times the action of a soft current as an operator in color space on the latter. The corresponding universal factorization limit has been known for a long time at the tree level \cite{Weinberg:1965nx}. At the one-loop level the soft limit was first extracted from color-ordered amplitudes \cite{Bern:1995ix,Bern:1998sc,Bern:1999ry} before the full color operator structure of the soft current was uncovered \cite{Catani:2000pi}. The two-loop soft current was extracted to finite order in the dimensional regulator~$\epsilon$ in ref.~\cite{Badger:2004uk} by taking the soft limit of the two-loop splitting function. These results were extended to higher orders in $\epsilon$ in ref.~\cite{Li:2013lsa,Duhr:2013msa}, allowing the two-loop soft current to be used in the calculation of the threshold approximation of the N$^3$LO Higgs boson cross section \cite{Anastasiou:2014vaa}.  The two-loop order is also the first order where the soft limit can lead to color correlations of three hard partons, and the calculation of the corresponding current was presented in ref.~\cite{Dixon:2019lnw}. Beyond the single-soft emission current, the double-soft current has also been known at tree level for quite some time \cite{Catani:1999ss} and more recently also at the one-loop level \cite{Zhu:2020ftr,Catani:2022hkb,Czakon:2022dwk}. Finally, the triple soft limit is known at tree-level \cite{Catani:2019nqv,DelDuca:2022noh}.

The main methods used so far for calculating soft currents have been either extractions from amplitude calculations, or direct calculations via the Wilson line/SCET formalism. In this work, we introduce a new infrared subgraph expansion approach which we employ directly at the integrand level of the full amplitude.
Using this new technique we circumvent the very challenging task of computing a full three-loop scattering amplitude.
An infrared-subgraph-finding algorithm, which can be seen as a generalization of the expansion-by-subgraph approach \cite{Smirnov:1990rz,Smirnov:1994tg} from Euclidean space to Minkowski space, was recently developed in the context of on-shell expansions for wide-angle scattering \cite{Gardi:2022khw}. Here we outline how to employ the same algorithm to find the set of infrared subgraphs contributing to the soft expansion. We present a general strategy for the soft expansion at arbitrary loop order with emphasis on the single-soft case. In particular, we provide criteria to identify infrared subgraphs which are in one-to-one correspondence with regions identified by the method-of-regions approach in parametric space, as implemented  in \cite{PakSmn11,Jantzen:2012mw,Ananthanarayan:2018tog,SmnvSmnSmv19,Heinrich:2021dbf}. The calculation of the three-loop soft current not only represents a new result, but also serves as a proof of concept demonstrating the potential of the expansion-by-subgraph approach in a highly non-trivial example. Indeed, this approach has been employed before in next-to-soft expansions of the Drell-Yan cross section at NNLO \cite{Bonocore:2014wua}, as well as at higher orders in the soft expansion in double-real-virtual and real-virtual squared corrections to the N$^3$LO Higgs boson cross section  \cite{Anastasiou:2013mca,Anastasiou:2015yha,Anastasiou:2013srw}; however this marks the first application of a fully systematic and automated approach in momentum space.

Our approach facilitates the generation of an integrand in the soft limit of scattering amplitudes. The challenging task of actually performing the loop integration remains. 
To this end, we employ integration-by-parts (IBP) identities~\cite{Laporta:2001dd,Chetyrkin1981,Tkachov1981} to express our amplitudes in terms of soft master integrals.
We then introduce another scale into these soft MIs by completing the square of a certain subset of propagators, which are linear in the loop momentum. The resulting integrals can be identified as \emph{collinear} MIs and contain the original soft MIs in their soft limits. We solve these collinear MIs via the method of differential equations~\cite{Henn:2013pwa,Gehrmann:1999as,Kotikov:1990kg,Kotikov:1991hm,Kotikov:1991pm} in terms of harmonic polylogarithms~\cite{Remiddi:1999ew,Maitre:2005uu} up to weight eight. Apart from one simple integral, we find that the soft boundary integrals are fully determined from regularity and consistency conditions~\cite{Henn:2020lye,Dulat:2014mda,Henn:2013nsa} of the system of differential equations.

The main result of this article is the extraction of three-loop QCD corrections to the single-soft emission current acting on two additional colored partons. 
In addition, we use our techniques to perform a computation of the single-soft limit of the stress tensor multiplet three-point form factor in $\mathcal{N}=4$ sYM theory based on an integrand provided in ref.~\cite{Lin:2021kht}.
Our calculation explicitly confirms the principle of maximal transcendentality~\cite{Kotikov:2004er,Kotikov:2002ab} for the contribution to the single-soft emission current considered in this article. 
Furthermore, we use our newly obtained results for the soft current at three loops to derive contributions to the Higgs boson and Drell-Yan production cross section at N$^4$LO in perturbative QCD due to single real emission contributions in the soft limit.

The remainder of this work is organized as follows. In section \ref{sec:setup} we introduce notation and the main results for the three-loop single-soft current with two colored partons. We describe the steps of our calculation in section~\ref{sec:calc}. In section~\ref{sec:regions} we provide a detailed description of our new integrand expansion technique. In section~\ref{sec:IR} we discuss the universal pole structure of the soft limit of our newly computed scattering amplitudes as a consistency check. Next, we discuss results for the three-loop form factor in $\mathcal{N}=4$ sYM theory in section~\ref{sec:N4SYM}. Furthermore, we present the threshold limit to single-real emission contributions to the Higgs boson and Drell-Yan production cross section at threshold at N$^4$LO in QCD perturbation theory in section~\ref{sec:thresholdxs}. Finally, we conclude in section~\ref{sec:conclusions}.

%% file: Chapters/SetUp.tex
\section{Single-soft current up to three loops in QCD}
\label{sec:setup}

We consider a scattering amplitude $\mathcal{A}$ in which the momentum $q$ of a single gluon is very low-energetic (i.e. the gluon is soft). In this single-soft limit, scattering amplitudes factorize into a universal operator acting on the scattering amplitude without the soft gluon ~\cite{Weinberg:1965nx,Yennie:1961ad}.
\beq
\label{eq:softfac}
\lim_{q \to 0} \mathcal{A}_{p_1 p_2 \dots p_n q} = {\mathbf J(q)} \mathcal{A}_{p_1 p_2 \dots p_n }.
\eeq
The operator $\mathbf{J}$ is referred to as the \emph{single-soft emission current} and acts on the colored degrees of freedom of the scattering amplitude $\mathcal{A}_{p_1 p_2 \dots p_n }$. In general, this operator will correlate emissions from all color-charged particles of the scattering amplitude. In this article, we determine the contribution to the soft-emission current that correlates two color-charged particles through three loops in perturbative QCD and $\mathcal{N}=4$ sYM theory. Our result is exact for scattering amplitudes involving only two color-charged external particles on top of the emitted soft gluon. 
Furthermore, our results represent an important first step in determining $\mathbf{J}(q)$ to third order in the coupling constant.

Up to three-loop order, the single-soft emission current can be decomposed as follows.
\bea
\label{eq:currentstruc}
 {\mathbf J(q)}&=&  \frac{i g_S}{C_A}  \epsilon^{a}_\mu (q) \sum_{i\neq j}   \left(\frac{p_i^\mu}{p_i\cdot q}-\frac{p_j^\mu}{p_j\cdot q}\right)  \left[ f^{abc}{\mathbf T_i^b}{\mathbf T_j^c}K_{2}(q,p_i,p_j) \right.\nonumber\\
 &&\left.+i C_A\left(d_{4A}^{abcd}K_{4A}(q,p_i,p_j)+n_f d_{4F}^{abcd}K_{4F}(q,p_i,p_j)\right)  \left\{ {\mathbf T_i^b},{\mathbf T_i^c} \right\} {\mathbf T_j^d}\right].
\eea
The bold faced notation above indicates an operator acting on the color structure of the amplitude. Note that the general structure of the ${\mathbf J(q)}$ can be more complex if its action on amplitudes with more than two colored particles is considered~\cite{Dixon:2019lnw}.
We work using dimensional regularization as a framework to regulate ultraviolet and infrared singularities, with $\epsilon$ as the dimensional regulator related to the spacetime dimension via $d=4-2\epsilon$. The number of quark flavors is given by $n_f$. The index $i$ sums over all color-charged particles of the scattering amplitude (to which the current is applied).
The factors $K_X$ are scalar quantities that can be expanded in perturbative QCD as follows.
\beq
\label{eq:coefdef}
K_X(q,p_i,p_j) =\sum_{o=0}^\infty a_S^o \left(\frac{(-2qp_i-i 0)(-2qp_j-i 0)}{(-2p_ip_j-i0) \mu^2}\right)^{-o \epsilon}K_X^{(o)}.
\eeq
The scalar products of the momenta appearing above are equipped with an infinitesimal imaginary part, inherited from Feynman's $i0$ prescription. It is our convention that all momenta are treated as incoming. Consequently, all scalar products are positive such that the term in the bracket in eq.~(\ref{eq:coefdef}) above introduces imaginary parts to the scattering amplitude. 
If one computes other configurations of scattering of particles (incoming and outgoing), then the corresponding soft-emission current can be derived by the appropriate crossing and analytic continuation according to the $i0$ prescription indicated earlier.
Above, $a_S$ is related to the bare strong coupling constant $\alpha_S^0$ by some universal factors.
\beq
\label{eq:asdef}
a_S=\frac{\alpha_S^0}{\pi} \left(\frac{4\pi}{ \mu^2}\right)^{\epsilon} e^{-\gamma_E \epsilon},\hspace{1cm}\gamma_E=0.577216\dots .
\eeq
The coefficients $K^{(o)}_X$ have been computed in perturbative QCD at one loop~\cite{Catani:2000pi,Bern:1995ix,Bern:1998sc,Bern:1999ry} and two loops~\cite{Badger:2004uk,Duhr:2013msa,Li:2013lsa,Dixon:2019lnw}, where only the terms $K_2^{(o)}$ are non-zero. From three loops, non-vanishing contributions from $K_{4A}^{(3)}$ and $K_{4F}^{(3)}$ emerge. The color tensors $d_{4A}^{abcd}$ and their contractions are defined as follows:
\beq
C_4^{R_1R_2} = d_{R_1}^{abcd} d_{R_2}^{abcd},\hspace{1cm}d_R^{abcd}= \frac{1}{4!} \left[\text{Tr}\big(T_R^aT_R^bT_R^cT_R^d\big) +\text{symmetric permutations}\right]\,.
\eeq
Above $T_R^a$ are the generators in representation $R$ of a general compact semi-simple Lie algebra; explicitly for the fundamental and adjoint representation we find:
\beq
T_{F,\,ij}^a =T^a_{ij}\,,\hspace{1cm} T_{A,\, ij}^a=-if^{aij}.
\eeq
The labels $A$ and $F$ refer to the adjoint and fundamental representations respectively. For $SU(n_c)$ we can express the quartic Casimirs in terms of the number of colors~$n_c$:
\beq
C_4^{AA}=\frac{n_c^2}{24} (n_c^2-1)(36+n_c^2),\quad
C_4^{AF}=\frac{n_c}{48} (n_c^2-1)(6+n_c^2),\quad
C_4^{FF}=\frac{(n_c^2-1)(18-6n_c^2+n_c^4)}{96n_c^2} .
\eeq
One of the main results of this article is the explicit computation of the coefficients $K_X^{(o)}$ through three-loop order. These coefficients are stated here explicitly and are also provided as ancillary files to the arXiv submission of this article. Their computation will be discussed in more detail later.
\bea
K_2^{(0)}&=&1.\\
K_2^{(1)}&=&-C_A\frac{e^{\gamma \epsilon } \Gamma (1-\epsilon )^3 \Gamma (\epsilon +1)^2}{4 \epsilon ^2 \Gamma (1-2 \epsilon )} \\
&=&C_A \Bigg[
-\frac{1}{4 \epsilon ^2}-\frac{\zeta _2}{8}+\frac{7 \zeta _3 \epsilon }{12}+\frac{39 \zeta_4 \epsilon ^2}{64}+\left(\frac{7 \zeta _2 \zeta _3}{24}+\frac{31 \zeta _5}{20}\right) \epsilon ^3+\left(\frac{1555 \zeta_6}{512}-\frac{49 \zeta _3^2}{72}\right) \epsilon ^4\nonumber\\
&&+\left(-\frac{91}{64} \zeta _3 \zeta _4+\frac{31 \zeta _2 \zeta _5}{40}+\frac{127 \zeta _7}{28}\right) \epsilon ^5+\left(-\frac{49}{144} \zeta _2 \zeta _3^2-\frac{217 \zeta _5 \zeta _3}{60}+\frac{37009 \zeta _8}{4096}\right) \epsilon^6 \Bigg]+\mathcal{O}(\epsilon^7).\nonumber
\eea

\bea
K_2^{(2)}&=&C_A^2 \Bigg[
\frac{1}{32 \epsilon ^4}
-\frac{11}{192 \epsilon ^3}
+\left(\frac{\zeta _2}{16}
-\frac{67}{576}\right)\frac{1}{\epsilon ^2}
+\left(-\frac{11 \zeta _2}{192}-\frac{11 \zeta _3}{96}-\frac{193}{864}\right)\frac{1}{\epsilon } \\ 
&&-\frac{67 \zeta _2}{576}+\frac{341 \zeta _3}{288}+\frac{7 \zeta _4}{128}-\frac{571}{1296}\nonumber \\ 
&&+\left(-\frac{7}{96} \zeta _3 \zeta _2-\frac{139 \zeta _2}{864}+\frac{2077 \zeta _3}{864}+\frac{2035 \zeta _4}{768}-\frac{247 \zeta _5}{160}-\frac{1705}{1944}\right) \epsilon\nonumber \\ 
&&+\left(-\frac{205 \zeta _3^2}{288}+\frac{341 \zeta _2 \zeta _3}{288}+\frac{1597 \zeta _3}{324}-\frac{109 \zeta _2}{324}+\frac{12395 \zeta _4}{2304}+\frac{5621 \zeta _5}{480}-\frac{3307 \zeta _6}{768}-\frac{5107}{2916}\right) \epsilon ^2\nonumber \\ 
&&+\left(-\frac{10571 \zeta _3^2}{864}+\frac{2077 \zeta _2 \zeta _3}{864}-\frac{509 \zeta _4 \zeta _3}{384}+\frac{37427 \zeta _3}{3888}-\frac{2411 \zeta _2}{3888}+\frac{41105 \zeta _4}{3456}-\frac{219 \zeta _2 \zeta _5}{160}\right.\nonumber\\
&&\left.+\frac{34237 \zeta _5}{1440}+\frac{42361 \zeta _6}{1024}-\frac{4573 \zeta _7}{224}-\frac{15313}{4374}\right) \epsilon ^3\nonumber \\ 
&&+\epsilon ^4 \left(-\frac{5 \zeta _{5,3}}{2}-\frac{845}{288} \zeta _2 \zeta _3^2-\frac{64387 \zeta _3^2}{2592}+\frac{1381 \zeta _2 \zeta _3}{324}-\frac{63085 \zeta _4 \zeta _3}{1152}-\frac{29 \zeta _5 \zeta _3}{240}\right.\nonumber\\
&&\left.+\frac{226405 \zeta _3}{11664}-\frac{14785 \zeta _2}{11664}+\frac{119135 \zeta _4}{5184}+\frac{5621 \zeta _2 \zeta _5}{480}+\frac{27187 \zeta _5}{540}+\frac{258017 \zeta _6}{3072}+\frac{90101 \zeta _7}{672}\right.\nonumber\\
&&\left.-\frac{1264777 \zeta _8}{18432}-\frac{45931}{6561}\right)\Bigg]\nonumber \\ 
&&+n_f C_A\Bigg[ \frac{1}{96 \epsilon ^3} 
+\frac{5}{288 \epsilon ^2}
+\left(\frac{\zeta _2}{96}+\frac{19}{864}\right)\frac{1}{\epsilon } 
+\frac{5 \zeta _2}{288}-\frac{31 \zeta _3}{144}+\frac{65}{2592}\nonumber \\ 
&&+\left(-\frac{35 \zeta _2}{864}-\frac{155 \zeta _3}{432}-\frac{185 \zeta _4}{384}+\frac{211}{7776}\right) \epsilon\nonumber \\ 
&&+\left(-\frac{31}{144} \zeta _3 \zeta _2-\frac{367 \zeta _2}{2592}-\frac{497 \zeta _3}{648}-\frac{925 \zeta _4}{1152}-\frac{511 \zeta _5}{240}+\frac{665}{23328}\right) \epsilon ^2\nonumber \\ 
&&+\left(\frac{961 \zeta _3^2}{432}-\frac{155 \zeta _2 \zeta _3}{432}-\frac{5255 \zeta _3}{3888}-\frac{3083 \zeta _2}{7776}-\frac{8915 \zeta _4}{3456}-\frac{511 \zeta _5}{144}-\frac{3851 \zeta _6}{512}+\frac{2059}{69984}\right) \epsilon ^3\nonumber \\ 
&&+\left(\frac{4805 \zeta _3^2}{1296}-\frac{65 \zeta _2 \zeta _3}{648}+\frac{5735 \zeta _4 \zeta _3}{576}-\frac{15623 \zeta _3}{5832}-\frac{20503 \zeta _2}{23328}-\frac{55225 \zeta _4}{10368}-\frac{511 \zeta _2 \zeta _5}{240}\right.\nonumber\\
&&\left.-\frac{9917 \zeta _5}{1080}-\frac{19255 \zeta _6}{1536}-\frac{8191 \zeta _7}{336}+\frac{6305}{209952}\right) \epsilon ^4 \Bigg]+ \mathcal{O}(\epsilon^5). \nonumber
\eea

\bea
\label{eq:K23}
K_2^{(3)}&=&C_A^3\Bigg[-\frac{1}{384 \epsilon ^6}
+\frac{11}{768 \epsilon ^5} 
+\left(\frac{119}{20736}-\frac{3 \zeta _2}{256}\right)\frac{1}{\epsilon ^4} 
+\left(\frac{649 \zeta _2}{13824}+\frac{\zeta _3}{96}-\frac{1517}{31104}\right)\frac{1}{\epsilon ^3} \\ 
&&+\left(\frac{2501 \zeta _2}{41472}-\frac{2101 \zeta _3}{6912}-\frac{1487 \zeta _4}{18432}-\frac{7271}{31104}\right)\frac{1}{\epsilon ^2} \nonumber\\
&&+\left(\frac{11 \zeta _3 \zeta _2}{1152}+\frac{437 \zeta _2}{62208}+\frac{2575 \zeta _3}{2304}-\frac{22583 \zeta _4}{36864}+\frac{49 \zeta _5}{160}-\frac{446705}{559872}\right)\frac{1}{\epsilon}+\frac{293 \zeta _3^2}{2304} \nonumber \\ 
&&-\frac{2453 \zeta _2 \zeta _3}{4608}+\frac{203705 \zeta _3}{31104}-\frac{12911 \zeta _2}{186624}+\frac{493381 \zeta _4}{110592}-\frac{26543 \zeta _5}{3840}+\frac{445679 \zeta _6}{442368}-\frac{8206861}{3359232}\nonumber \\ 
&&+\left(-\frac{17149 \zeta _3^2}{13824}+\frac{21031 \zeta _2 \zeta _3}{13824}+\frac{43 \zeta _4 \zeta _3}{288}+\frac{2330483 \zeta _3}{93312}-\frac{403379 \zeta _2}{1119744}+\frac{1228523 \zeta _4}{55296}\right.\nonumber\\
&&\left.+\frac{9773 \zeta _2 \zeta _5}{5760}+\frac{262597 \zeta _5}{11520}-\frac{25965643 \zeta _6}{884736}+\frac{151631 \zeta _7}{16128}-\frac{48027739}{6718464}\right) \epsilon\nonumber \\ 
&&+ \left(-\frac{469 \zeta _{5,3}}{90}+\frac{10045 \zeta _2 \zeta _3^2}{4608}-\frac{920995 \zeta _3^2}{13824}+\frac{71831 \zeta _2 \zeta _3}{6912}+\frac{388289 \zeta _4 \zeta _3}{36864}-\frac{9907 \zeta _5 \zeta _3}{1920}\right.\nonumber\\
&&\left.+\frac{15854467 \zeta _3}{186624}-\frac{5363867 \zeta _2}{6718464}+\frac{42678481 \zeta _4}{497664}-\frac{71533 \zeta _2 \zeta _5}{7680}+\frac{82837 \zeta _5}{640}\right.\nonumber\\
&&\left.+\frac{112195243 \zeta _6}{884736}-\frac{1343045 \zeta _7}{8064}+\frac{3738034847 \zeta _8}{53084160}-\frac{2482106477}{120932352}\right)\epsilon ^2 \Bigg] \nonumber \\ 
&+&C_A^2 n_f \Bigg[
-\frac{1}{384 \epsilon ^5}
+\frac{43}{10368 \epsilon ^4} 
+\left(\frac{895}{31104}-\frac{59 \zeta _2}{6912}\right)\frac{1}{\epsilon ^3}
+\left(-\frac{31 \zeta _2}{20736}+\frac{239 \zeta _3}{3456}+\frac{2603}{31104}\right)\frac{1}{\epsilon ^2}\nonumber \\ 
&&+\left(\frac{3265 \zeta _2}{62208}-\frac{4945 \zeta _3}{10368}+\frac{2437 \zeta _4}{18432}+\frac{24169}{139968}\right)\frac{1}{\epsilon }\nonumber \\ 
&&+\frac{271 \zeta _3 \zeta _2}{2304}-\frac{3925 \zeta _2}{186624}-\frac{2513 \zeta _3}{1152}-\frac{33109 \zeta _4}{18432}+\frac{7799 \zeta _5}{5760}+\frac{397699}{1679616}\nonumber \\ 
&&+\left(-\frac{4969 \zeta _3^2}{6912}-\frac{1595 \zeta _2 \zeta _3}{2304}-\frac{720299 \zeta _3}{93312}-\frac{228895 \zeta _2}{279936}-\frac{1168171 \zeta _4}{165888}-\frac{187753 \zeta _5}{17280}+\frac{2476865 \zeta _6}{442368}\right.\nonumber\\
&&\left.-\frac{22273}{373248}\right) \epsilon 
+\left(\frac{404075 \zeta _3^2}{20736}-\frac{78295 \zeta _2 \zeta _3}{20736}-\frac{121555 \zeta _4 \zeta _3}{18432}-\frac{3316207 \zeta _3}{139968}-\frac{17477627 \zeta _2}{3359232}\right.\nonumber\\
&&\left.-\frac{15232813 \zeta _4}{497664}+\frac{7063 \zeta _2 \zeta _5}{3840}-\frac{52115 \zeta _5}{1152}-\frac{76597939 \zeta _6}{1327104}+\frac{13871 \zeta _7}{448}-\frac{125652667}{60466176}\right) \epsilon ^2 \Bigg] \nonumber\\
&+&C_F C_A n_f\Bigg[
\frac{1}{576 \epsilon ^3} 
+\left(\frac{55}{3456}-\frac{\zeta _3}{72}\right)\frac{1}{\epsilon ^2}
+\left(\frac{\zeta _2}{384}-\frac{19 \zeta _3}{432}-\frac{\zeta _4}{48}+\frac{1819}{20736}\right)\frac{1}{\epsilon }\nonumber \\ 
&&-\frac{1}{48} \zeta _3 \zeta _2+\frac{67 \zeta _2}{2304}-\frac{1385 \zeta _3}{5184}-\frac{19 \zeta _4}{288}-\frac{7 \zeta _5}{72}+\frac{45967}{124416}\nonumber \\ 
&&+\left(\frac{17 \zeta _3^2}{18}-\frac{19 \zeta _2 \zeta _3}{288}-\frac{50495 \zeta _3}{31104}+\frac{3547 \zeta _2}{13824}-\frac{16237 \zeta _4}{27648}-\frac{133 \zeta _5}{432}-\frac{101 \zeta _6}{384}+\frac{1007179}{746496}\right) \epsilon\nonumber \\ 
&&+\left(\frac{323 \zeta _3^2}{108}-\frac{809 \zeta _2 \zeta _3}{3456}+\frac{599 \zeta _4 \zeta _3}{128}-\frac{1661303 \zeta _3}{186624}+\frac{99931 \zeta _2}{82944}-\frac{635899 \zeta _4}{165888}\right.\nonumber\\
&&+\left.-\frac{7 \zeta _2 \zeta _5}{48}-\frac{70417 \zeta _5}{25920}-\frac{1919 \zeta _6}{2304}-\frac{49 \zeta _7}{72}+\frac{20357263}{4478976}\right) \epsilon ^2
\Bigg]+\mathcal{O}(\epsilon^3)\nonumber\\
&+&C_A n_f^2 \Bigg[
-\frac{1}{1296 \epsilon ^4} 
-\frac{5}{1944 \epsilon ^3} 
+\left(-\frac{\zeta _2}{864}-\frac{1}{216}\right)\frac{1}{\epsilon ^2} 
+\left(-\frac{5 \zeta _2}{1296}+\frac{65 \zeta _3}{1296}-\frac{11}{2187}\right)\frac{1}{\epsilon }\nonumber \\ 
&&+\frac{11 \zeta _2}{432}+\frac{325 \zeta _3}{1944}+\frac{1229 \zeta _4}{6912}+\frac{10}{6561}
+\Bigg(\frac{65 \zeta _3 \zeta _2}{864}+\frac{187 \zeta _2}{1458}+\frac{37 \zeta _3}{72}+\frac{6145 \zeta _4}{10368}\nonumber \\
&&+\frac{2521 \zeta _5}{2160}+\frac{190}{6561}\Bigg) \epsilon
+\Bigg(-\frac{4225 \zeta _3^2}{2592}+\frac{325 \zeta _2 \zeta _3}{1296}+\frac{2632 \zeta _3}{2187}+\frac{1058 \zeta _2}{2187}
+\frac{9355 \zeta _4}{3456}\nonumber \\ 
&&
+\frac{2521 \zeta _5}{648}+\frac{999593 \zeta _6}{165888}+\frac{6614}{59049}\Bigg) \epsilon ^2 \Bigg]\nonumber.
\eea

\bea
K_{4A}^{(0)}&=&K_{4A}^{(1)}=K_{4A}^{(2)}=0.\\
K_{4A}^{(3)}
\label{eq:K4A3}
&=&
\left(\frac{\zeta _2 \zeta _3}{8}+\frac{\zeta _5}{16}\right)\frac{1}{\epsilon}
+\frac{3 \zeta _3^2}{4}-\frac{\zeta _3}{12}+\frac{\zeta _2}{4}-\frac{55 \zeta _5}{24}+\frac{235 \zeta _6}{64} \\ 
&&+\left(-\frac{77 \zeta _3^2}{12}+\frac{139 \zeta _4 \zeta _3}{32}+\frac{53 \zeta _3}{72}+\frac{13 \zeta _2}{8}-\frac{13 \zeta _4}{16}+\frac{187 \zeta _2 \zeta _5}{32}-\frac{335 \zeta _5}{72}-\frac{55 \zeta _6}{8}+\frac{63 \zeta _7}{4}\right) \epsilon\nonumber \\ 
&&+\left(-\frac{459 \zeta _{5,3}}{20}+\frac{15}{8} \zeta _2 \zeta _3^2-\frac{469 \zeta _3^2}{36}-\frac{19 \zeta _2 \zeta _3}{8}-\frac{77 \zeta _4 \zeta _3}{4}-\frac{343 \zeta _5 \zeta _3}{16}+\frac{665 \zeta _3}{108}\right. \nonumber\\
&&\left.+\frac{97 \zeta _2}{12}+\frac{157 \zeta _4}{12}-\frac{55 \zeta _2 \zeta _5}{16}-\frac{3163 \zeta _5}{216}-\frac{335 \zeta _6}{24}-\frac{1705 \zeta _7}{16}+\frac{1306649 \zeta _8}{5760}\right)\epsilon ^2 
+\mathcal{O}(\epsilon^3)\nonumber.
\eea

\bea
K_{4F}^{(0)}&=&K_{4F}^{(1)}=K_{4F}^{(2)}=0.\\
K_{4F}^{(3)}
\label{eq:K4F3}
&=&
-\frac{\zeta _2}{2}+\frac{\zeta _3}{6}+\frac{5 \zeta _5}{6} + \left(\frac{7 \zeta _3^2}{3}-\frac{47 \zeta _3}{36}-\frac{15 \zeta _2}{4}+\frac{13 \zeta _4}{8}+\frac{25 \zeta _5}{18}+\frac{5 \zeta _6}{2}\right) \epsilon \\ 
&&\left(\frac{35 \zeta _3^2}{9}+\frac{19 \zeta _2 \zeta _3}{4}+7 \zeta _4 \zeta _3-\frac{1471 \zeta _3}{108}-\frac{239 \zeta _2}{12}-\frac{589 \zeta _4}{24}+\frac{5 \zeta _2 \zeta _5}{4}+\frac{1423 \zeta _5}{108}\nonumber \right. \\
&&\left.+\frac{25 \zeta _6}{6}+\frac{155 \zeta _7}{4}\right) \epsilon ^2+\mathcal{O}(\epsilon^3).\nonumber
\eea

%% file: Chapters/Computation.tex
\section{Calculation of the soft limit of scattering amplitudes}
\label{sec:calc}

In order to compute three-loop corrections to the single-soft emission current, we calculate the soft limit of physical scattering amplitudes.
In particular, we compute the limit of the scattering amplitude involving a Higgs boson and three gluons, as well as the limit of the amplitude involving an off-shell transverse photon, a gluon, and a quark-antiquark pair. 
These scattering amplitudes at three-loop order are contributions to the production cross section of a Higgs boson or vector boson in association with a jet at the LHC at N$^3$LO in perturbative QCD, for example. 
For the purposes of our calculation, we work in a kinematic configuration where the color singlet boson (with momentum $p_1$) decays to three color charged partons (with momenta $p_2$, $p_3$, and $p_4$).
\beq
\label{eq:p1_to_p2p3p4_decay}
h (p_1)\to g(p_2)\, g(p_3)\, g(p_4) ,\hspace{1cm} \gamma^*(p_1)\to q(p_2)\, \bar q(p_3) \, g(p_4).
\eeq
These scattering amplitudes were computed through two-loop order in refs.~\cite{Gehrmann:2023zpz,Gehrmann:2023etk,Gehrmann:2022vuk,Gehrmann:2011aa,Garland:2001tf,Garland:2002ak}, and first results for planar contributions to the virtual photon amplitudes have appeared recently in ref.~\cite{Gehrmann:2023jyv}. 
However, a complete set of three-loop amplitudes is still elusive. 
In the case of Higgs boson scattering amplitudes, we work in the limit where the top quark is treated as infinitely massive and its degrees of freedom are integrated out~\cite{Spiridonov:1988md,Inami1983,Shifman1978,Wilczek1977}. 
As a result, a dimension-five operator is introduced that couples~\cite{Chetyrkin:2005ia,Schroder:2005hy,Chetyrkin:1997un,Kramer:1996iq} the Higgs boson directly to the gluon field strength. We treat all quarks as massless and work with $n_f$ light quark degrees of freedom.

We create the integrands for our desired scattering amplitudes by generating Feynman diagrams using the QGRAF program~\cite{qgraf} and dressing them with QCD Feynman rules. To facilitate our computation, we define suitable projectors to Lorentz tensor structures whose scalar coefficients we compute in the so-called conventional dimensional regularization (CDR) scheme employing standard methods.

Once we obtain scalar integrals, it is our goal to avoid the complexity of computing full, three-loop scattering amplitudes. 
Instead, we develop a new technique, based on the method of regions~\cite{Beneke:1997zp}, to expand the scalar integrands around the limit of the gluon momentum $p_4$ becoming soft (i.e. $p_4\to 0$).
We identify that the contribution to the single-soft emission current is provided by the \emph{maximally soft regions} of our integrals.
In these regions, all the loop momenta are equally as soft as the external soft gluon with momentum $p_4$. We keep only the first term in the expansion for both the Higgs boson and virtual photon amplitudes.
More details regarding this expansion technique will be discussed in section~\ref{sec:regions}.

Once an expanded scalar integrand is obtained, we use standard multi-loop techniques in order to integrate over all the loop momenta.
First, we use integration-by-part (IBP) identities~\cite{Laporta:2001dd,Chetyrkin1981,Tkachov1981} in the form of the Laporta algorithm~\cite{Laporta:2001dd} to relate all the soft Feynman integrals to a set of soft master integrals. 
These soft master integrals only depend on the external kinematics via an overall multiplicative prefactor~\cite{Zhu:2014fma}. 
For example, one soft integral contributing at two-loop order is given by 
\bea
I&=&\int \frac{d ^d p_5}{(2\pi)^d}\frac{d^d p_6}{(2\pi)^d}\frac{1}{[p_5^2][(p_5-p_6)^2][(p_5-p_6)^2][(p_5-p_6-p_4)^2]}\\
&\times&\frac{1}{[2p_2 p_5][-2p_3 p_6][-2p_3 p_5][2p_2 p_4+2p_2p_6]}.\nonumber
\eea
All propagators involving the hard external momenta $p_2$ and $p_3$ were linearized in the expansion procedure. Consequently, it is now possible to read off the integer power of the dependence of the integral on $p_2$ and $p_3$ directly from the integrand (see for example refs.~\cite{Duhr:2022cob,Anastasiou:2015yha} for details).
It is exactly this property, in addition to knowing the overall energy dimension of the integral, that fixes all kinematic dependence of the integral and determines it up to a function in the space-time dimension~$d$:
\beq
I=(2p_2p_3)^{-1+2\epsilon}(2p_2p_4)^{-1-2\epsilon}(2p_3p_4)^{-1-2\epsilon} F(\epsilon).
\eeq

Especially at three-loop order, computing the remaining soft master integrals using straightforward integration techniques is challenging.
Thus, we follow a different path and temporarily undo our soft expansion for all those propagators depending on one of the hard external momenta, $p_3$.
\beq
\frac{1}{[-2p_3 p_6][-2p_3 p_5]}\ \to\ 
\frac{1}{[ (p_6-p_3)^2][(p_5-p_3)^2]}.
\eeq
It is now no longer possible to read off the dependence of the integral on $p_3$ from the integrand, and the result will consequently be a nontrivial function of the dimensionless ratio 
\beq
w=\frac{s_{24} }{s_{23}},\hspace{1cm} s_{ij}=(p_i+p_j)^2.
\eeq
We now apply the method of differential equations~\cite{Henn:2013pwa,Gehrmann:1999as,Kotikov:1990kg,Kotikov:1991hm,Kotikov:1991pm} to determine our integrals as a function of $w$. 
To accomplish this, we transform the differential equations into the canonical form~\cite{Henn:2013pwa} using algorithmic techniques~\cite{Lee:2014ioa}.
The solutions to these differential equations can be expressed in terms of harmonic polylogarithms in $w$~\cite{Remiddi:1999ew}. 
However, the differential equations determine the master integrals only up to boundary constants. 
To constrain them, we first compute differential equations for master integrals undoing our soft-expansion for propagators involving $p_2$ and $p_3$ separately. 
We then demand that the solutions are consistent among themselves when taking the strict soft limit $w\to 0$.
Demanding consistency relations from our system of differential equations, as in refs.~\cite{Henn:2020lye} and~\cite{Dulat:2014mda}, relates all the required boundary conditions to one soft master integral, which is easily computed using direct integration in parameter space.
Consequently, we determine all soft master integrals through transcendental weight eight.

The soft master integrals, which serve as building blocks for the computation of the single-soft emission current at three loops, are one of the main results of this article. 
In total, we compute 50 soft master integrals and label them with an index $i$:
\beq
M^i \equiv (4\pi)^{-3 \epsilon} e^{3\gamma_E \epsilon} \left(\frac{(-s_{24})(-s_{34})}{ (-s_{23})}\right)^{3\epsilon} \int \frac{d^d p_5}{(2\pi)^d} \frac{d^d p_6}{(2\pi)^d} \frac{d^d p_7}{(2\pi)^d}\ \mathcal{I}_i \Bigg|_{s_{23}=s_{24}=s_{34}=1}.
\eeq
Above, we set all kinematic Mandelstam invariants to unity and remove non-rational dependence on them via a universal prefactor. Furthermore, we anticipate the renormalization of the strong coupling constant and absorb some $\overline{\text{MS}}$ scheme prefactors. The integrand~$\mathcal{I}_i$ is a rational function of the Lorentz invariant scalar products of the internal loop momenta $\{p_5,p_6,p_7\}$ and the external parton momenta $\{p_2,p_3,p_4\}$. The soft integrals are related to canonical soft master integrals (i.e. functions of pure transcendental weight) by a linear transformation of the vector of soft master integrals. 
\beq
\label{eq:CanMsDef}
\vec{M}=T_{\text{can.}}(\epsilon) \cdot \vec{M}_c.
\eeq
The matrix $T_{\text{can.}}(\epsilon)$ only depends on the dimensional regulator and rational numbers. 
We provide the integrands $\mathcal{I}_i$, the transformation matrix $T_\text{can.}$, and solution to the canonical master integrals $M_c^i$ in ancillary files together with the arXiv submission of this article. 

Having calculated the strict soft limit of our scattering amplitudes, we can now extract the coefficients $K^{(o)}_X$ of eq.~\eqref{eq:coefdef} by identifying
\beq
\lim\limits_{\substack{\text{maximally soft}\\p_4\to0}} \mathcal{A}_{p_1 \to p_2p_3p_4} =\mathbf{J}(p_4)  \mathcal{A}_{p_1\to p_2 p_3}^{(0)}.
\eeq
Above, $\mathcal{A}^{(0)}$ is the tree-level scattering amplitude for the $2\to 1$ Born process not involving the soft gluon with momentum $p_4$.
We find complete agreement of our computation of the coefficients of the single-soft emission current from our $h\to ggg$ and $\gamma^* \to q\bar qg$ amplitude, which serves as a strong consistency check of our computation and of the color structure identified in eq.~\eqref{eq:currentstruc}.

%% file: Chapters/Regions.tex
\section{Regions in the soft expansion}
\label{sec:regions}
In this section, we develop a method for the expansion of scattering amplitudes in the limit of a set of partons becoming very low energetic (i.e. soft). 
The decay processes we introduced in eq.~(\ref{eq:p1_to_p2p3p4_decay}) is a specific case to which we apply this new technology. First, we introduce a general set-up which contains our expansion as a special case. Next, we explain how to identify the subgraphs which correspond to the regions of the expansion. Finally, we discuss the particular soft expansion of Feynman diagrams in our concrete setting.

To set up our expansion technique, we divide the external momenta of a Feynman graph into the following three categories:
 \begin{itemize}
    \item[(1)] $K$ massless momenta $p_i$,
    \item[(2)] $L$ off-shell momenta $q_j$,
    \item[(3)] $M$ soft massless momenta $l_k$.
 \end{itemize}
We define the soft expansion of a graph as an expansion around the limit $l_k\to 0$.
Scalar products involving of momenta including $l_k$ are consequently much smaller than scalar products not involving any $l_k$. Introducing a small parameter $\lambda$ and a hard reference scale~$Q$, we find
\begin{subequations}
\label{eq:wideangle_soft_gluon_kinematics}
\begin{align}
    & p_i^2=0\ \ (i=1,\dots,K), \quad q_j^2\sim Q^2\ \ (j=1,\dots,L), \quad l_k^2=0\ \ (k=1,\dots,M),\\
    & p_{i_1}\cdot p_{i_2}\sim Q^2\ \ (i_1\neq i_2), \quad p_i\cdot l_k\sim q_j\cdot l_k\sim \lambda Q^2, \quad l_{k_1}\cdot l_{k_2}\sim \lambda^2 Q^2\ \ (k_1\neq k_2).
\end{align}
\end{subequations}

\begin{figure}[t]
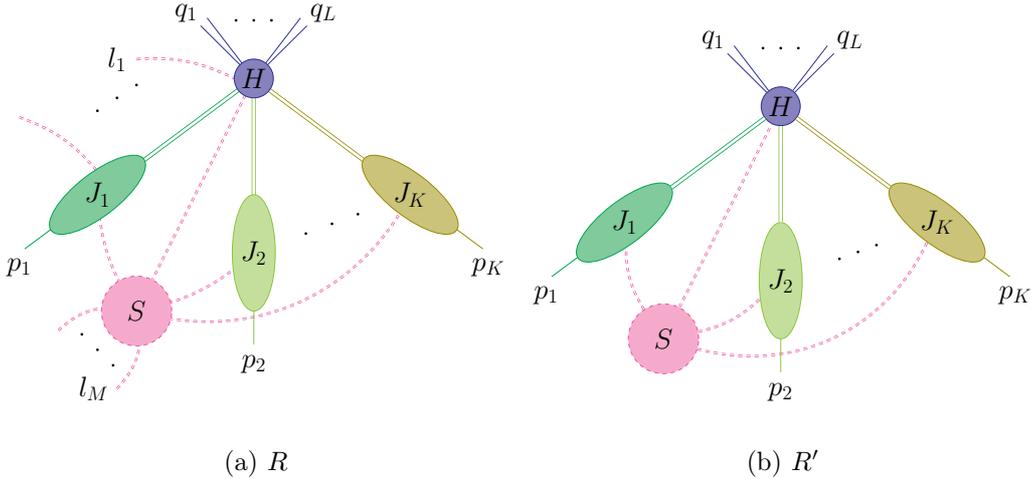

\centering
\begin{subfigure}[b]{0.45\textwidth}
    \centering
    \include{figs/soft_classical_picture}
    \caption{$R$}
    \label{soft_classical_picture}
\end{subfigure}
\begin{subfigure}[b]{0.45\textwidth}
    \centering
    \include{figs/onshell_classical_picture}
    \caption{$R'$}
    \label{onshell_classical_picture}
\end{subfigure}
\caption{The partitioning of a generic wide-angle scattering graph into infrared subgraphs corresponding to a particular region (a) $R$ in the soft expansion, eq.~(\ref{eq:wideangle_soft_gluon_kinematics}), and (b) $R'$ in the on-shell expansion, eq.~(\ref{eq:wideangle_onshell_kinematics}). The doubled lines connecting different blobs represent any number of propagators.}
\end{figure}

Our strategy of identifying the regions in the soft expansion is based on the observation that each region $R$ (see figure~\ref{soft_classical_picture}) must conform with the solutions of the Landau equations~\cite{Ma23}. Furthermore, once all the external soft momenta $l_k$ are removed from $R$, the resulting configuration $R'$ (see figure~\ref{onshell_classical_picture}) must be a region in the on-shell expansion developed in ref.~\cite{Gardi:2022khw}. In other words, the regions in the soft expansion can be derived from those in the on-shell expansion with additional requirements. The regions in the on-shell expansion have been studied in detail in ref.~\cite{Gardi:2022khw}, and in particular, a graph-finding algorithm was developed to obtain the complete list of regions from the given Feynman graph.
Here, we aim to leverage this knowledge to straightforwardly identify the regions in the soft expansion.
To this end, we will first review the graph-finding algorithm for the on-shell expansion and then delve into the generic configuration of regions in the soft expansion.

\subsection{The graph-finding algorithm for on-shell expansions}
\label{section-graph_finding_algorighm_onshell}

In the context of an on-shell expansion of wide-angle scattering, the Feynman graphs feature on-shell external momenta $p_1,\dots,p_K$ and off-shell external momenta $q_1,\dots, q_L$, satisfying
\begin{eqnarray}
\label{eq:wideangle_onshell_kinematics}
    p_i^2\sim \lambda Q^2\ \ (i=1,\dots,K),\quad q_j^2\sim Q^2\ \ (j=1,\dots,L),\quad p_{i_1}\cdot p_{i_2}\sim Q^2\ \ (i_1\neq i_2).
\end{eqnarray}
In contrast to the soft expansion defined in eq.~(\ref{eq:wideangle_soft_gluon_kinematics}), there are no soft external momenta present here, and every on-shell momentum $p_i$ is slightly off its corresponding lightcone.

A graph-finding algorithm has been provided in ref.~\cite{Gardi:2022khw}, along with a rigorous proof demonstrating that this algorithm generates all the regions for the on-shell expansion. This allows us to comprehend the structures of these regions and derive them more efficiently by circumventing the approach of constructing Newton polytopes in ref.~\cite{PakSmn11,Jantzen:2012mw,Ananthanarayan:2018tog,SmnvSmnSmv19}.

A key concept in this graph-finding algorithm is that of \emph{mojetic graphs}. We call a graph mojetic if it becomes \emph{one-vertex irreducible} after connecting all of its external edges to an auxiliary vertex. Note that a graph is called one-vertex irreducible if it remains connected after the removal of any one of its vertices. The algorithm can then be described by the following steps.
\begin{itemize}
    \item \emph{Step 1}: For each nearly on-shell external momentum $p_i$ ($i=1,\dots,K$), we draw a cut through a set of edges $\{e_c\}$ such that:
    \begin{enumerate}
        \item [1,] $\{e_c\}$ disconnects a graph $G$ into two connected subgraphs, one of which, denoted by $\widehat{\gamma}_i$, is attached by $p_i$ only;
        \item [2,] the graph $\gamma_i\equiv \widehat{\gamma}_i \cup \{e_c\}$ is mojetic.
    \end{enumerate}
    \item \emph{Step 2}: 
    For all possible sets $\{\gamma_1,\dots,\gamma_K\}$, we overlay the graphs $\gamma_1,\dots,\gamma_K$ and associate the edges $e\in G$ to certain subgraphs as follows. If~$e$ has been assigned to two or more $\gamma_{i}$, it belongs to the soft subgraph $S$; if~$e$ has been assigned to exactly one $\gamma_{i}$, it belongs to the jet subgraph $J_i$; if~$e$ has not been assigned to any $\gamma_{i}$, it belongs to $H$. Let us also denote $J\equiv \cup_{i=1}^K J_i$.
    \item \emph{Step 3}: We now require that the result obtained in \emph{Step 2} satisfies the following three further conditions: (i) each jet subgraph $J_i$ is connected; (ii) each hard subgraph $H$ is connected; (iii) each of the $K$ subgraphs $H \cup J \setminus J_i$ ($i=1,\dots,K$) is mojetic. The region would be ruled out if any of these conditions are not satisfied.
\end{itemize}
Let us illustrate how this algorithm works through the following example of a $3\times 2$ fishnet graph, which has four on-shell external momenta $p_1$, $p_2$, $p_3$ and $p_4$. A choice of the graphs $\gamma_1$, $\gamma_2$, $\gamma_3$ and $\gamma_4$, which satisfy the conditions outlined in \emph{Step 1}, is shown below. Note that in each figure, the edges $\{e_c\}$ are cut by the dotted curve.
\begin{equation}
\begin{aligned}
& \gamma_1:
\begin{tikzpicture}[baseline=5ex,scale=1.]
%\draw [help lines] (0,0) grid (5,4);
%coordinates:
\coordinate (x1) at (0,0) ;
\coordinate (x2) at (0,1) ;
\coordinate (x3) at (0,2) ;
\coordinate (x6) at (1,2) ;
\coordinate (x5) at (1,1) ;
\coordinate (x4) at (1,0) ;
\coordinate (x9) at (2,2) ;
\coordinate (x8) at (2,1) ;
\coordinate (x7) at (2,0) ;
\coordinate (x12) at (3,2) ;
\coordinate (x11) at (3,1) ;
\coordinate (x10) at (3,0) ;

%external momenta:
\node (q4) at (-0.5,-0.5) {$p_4$};
\node (p1) at (-0.5,2.5) {$p_1$};
\node (q3) at (3.5,-0.5) {$p_3$};
\node (p2) at (3.5,2.5) {$p_2$};

%soft:

%jets:4
\draw[color=green] (x1) -- (q4);

%1
\draw[color=LimeGreen] (x3) -- (p1);
\draw[color=LimeGreen] (x3) -- (x6);
\draw[color=LimeGreen] (x2) -- (x3);
\draw[color=LimeGreen] (x1) -- (x2);
\draw[color=LimeGreen] (x2) -- (x5);
\draw[color=LimeGreen] (x5) -- (x6);
\draw[color=LimeGreen] (x6) -- (x9);
\draw[color=LimeGreen] (x4) -- (x5);
\draw[color=LimeGreen] (x5) -- (x8);
\draw[fill,thick,color=LimeGreen] (x6) circle (1pt);
\draw[fill,thick,color=LimeGreen] (x2) circle (1pt);
\draw[fill,thick,color=LimeGreen] (x3) circle (1pt);
\draw[fill,thick,color=LimeGreen] (x5) circle (1pt);

%2
\draw[color=ForestGreen] (x12) -- (p2);

%3
\draw[color=Green] (x10) -- (q3);

%edges:

\draw[ultra thick,color=Blue] (x1) -- (x4);
%\draw[ultra thick,color=Blue] (x4) -- (x5);
\draw[ultra thick,color=Blue] (x4) -- (x7);
\draw[ultra thick,color=Blue] (x7) -- (x8);
\draw[ultra thick,color=Blue] (x8) -- (x9);
\draw[ultra thick,color=Blue] (x7) -- (x10);
\draw[ultra thick,color=Blue] (x8) -- (x11);
\draw[ultra thick,color=Blue] (x9) -- (x12);
\draw[ultra thick,color=Blue] (x10) -- (x11);
\draw[ultra thick,color=Blue] (x11) -- (x12);

%vertices:
\draw[fill,thick,color=Blue] (x1) circle (1pt);
\draw[fill,thick,color=Blue] (x4) circle (1pt);
\draw[fill,thick,color=Blue] (x7) circle (1pt);
\draw[fill,thick,color=Blue] (x8) circle (1pt);
\draw[fill,thick,color=Blue] (x9) circle (1pt);
\draw[fill,thick,color=Blue] (x10) circle (1pt);
\draw[fill,thick,color=Blue] (x11) circle (1pt);
\draw[fill,thick,color=Blue] (x12) circle (1pt);

\draw (-0.3,0.3) edge [dotted, ultra thick, color=Black, bend right =60] (1.7,2.3) node [right] {};

\end{tikzpicture}
\quad
\gamma_2:
\begin{tikzpicture}[baseline=5ex,scale=1.]
%\draw [help lines] (0,0) grid (5,4);
%coordinates:
\coordinate (x1) at (0,0) ;
\coordinate (x2) at (0,1) ;
\coordinate (x3) at (0,2) ;
\coordinate (x6) at (1,2) ;
\coordinate (x5) at (1,1) ;
\coordinate (x4) at (1,0) ;
\coordinate (x9) at (2,2) ;
\coordinate (x8) at (2,1) ;
\coordinate (x7) at (2,0) ;
\coordinate (x12) at (3,2) ;
\coordinate (x11) at (3,1) ;
\coordinate (x10) at (3,0) ;

%external momenta:
\node (q4) at (-0.5,-0.5) {$p_4$};
\node (p1) at (-0.5,2.5) {$p_1$};
\node (q3) at (3.5,-0.5) {$p_3$};
\node (p2) at (3.5,2.5) {$p_2$};

%jets:
\draw[color=green] (x1) -- (q4);
\draw[color=LimeGreen] (x3) -- (p1);
\draw[color=ForestGreen] (x12) -- (p2);
\draw[fill,thick,color=ForestGreen] (x8) circle (1pt);
\draw[fill,thick,color=ForestGreen] (x9) circle (1pt);
\draw[fill,thick,color=ForestGreen] (x11) circle (1pt);
\draw[fill,thick,color=ForestGreen] (x12) circle (1pt);
\draw[color=ForestGreen] (x5) -- (x8);
\draw[color=ForestGreen] (x6) -- (x9);
\draw[color=ForestGreen] (x7) -- (x8);
\draw[color=ForestGreen] (x8) -- (x9);
\draw[color=ForestGreen] (x8) -- (x11);
\draw[color=ForestGreen] (x9) -- (x12);
\draw[color=ForestGreen] (x10) -- (x11);
\draw[color=ForestGreen] (x11) -- (x12);

\draw[color=Green] (x10) -- (q3);

%edges:
\draw[ultra thick,color=Blue] (x1) -- (x2);
\draw[ultra thick,color=Blue] (x2) -- (x3);
\draw[ultra thick,color=Blue] (x1) -- (x4);
\draw[ultra thick,color=Blue] (x2) -- (x5);
\draw[ultra thick,color=Blue] (x3) -- (x6);
\draw[ultra thick,color=Blue] (x4) -- (x5);
\draw[ultra thick,color=Blue] (x5) -- (x6);
\draw[ultra thick,color=Blue] (x4) -- (x7);
\draw[ultra thick,color=Blue] (x7) -- (x10);

%vertices:
\draw[fill,thick,color=Blue] (x1) circle (1pt);
\draw[fill,thick,color=Blue] (x2) circle (1pt);
\draw[fill,thick,color=Blue] (x3) circle (1pt);
\draw[fill,thick,color=Blue] (x4) circle (1pt);
\draw[fill,thick,color=Blue] (x5) circle (1pt);
\draw[fill,thick,color=Blue] (x6) circle (1pt);
\draw[fill,thick,color=Blue] (x7) circle (1pt);
\draw[fill,thick,color=Blue] (x10) circle (1pt);

\draw (3.3,0.3) edge [dotted, ultra thick, color=Black, bend left =60] (1.3,2.3) node [right] {};

\end{tikzpicture}
\\
& \gamma_3:
\begin{tikzpicture}[baseline=5ex,scale=1.]
%\draw [help lines] (0,0) grid (5,4);
%coordinates:
\coordinate (x1) at (0,0) ;
\coordinate (x2) at (0,1) ;
\coordinate (x3) at (0,2) ;
\coordinate (x6) at (1,2) ;
\coordinate (x5) at (1,1) ;
\coordinate (x4) at (1,0) ;
\coordinate (x9) at (2,2) ;
\coordinate (x8) at (2,1) ;
\coordinate (x7) at (2,0) ;
\coordinate (x12) at (3,2) ;
\coordinate (x11) at (3,1) ;
\coordinate (x10) at (3,0) ;

%external momenta:
\node (p4) at (-0.5,-0.5) {$p_4$};
\node (p1) at (-0.5,2.5) {$p_1$};
\node (p3) at (3.5,-0.5) {$p_3$};
\node (p2) at (3.5,2.5) {$p_2$};

%jets:
\draw[color=green] (x1) -- (p4);
\draw[color=LimeGreen] (x3) -- (p1);
\draw[color=ForestGreen] (x12) -- (p2);
\draw[color=Green] (x10) -- (p3);
\draw[fill,thick,color=Green] (x10) circle (1pt);
\draw[color=Green] (x7) -- (x10);
\draw[color=Green] (x10) -- (x11);

%edges:
\draw[ultra thick,color=Blue] (x4) -- (x7);
\draw[ultra thick,color=Blue] (x1) -- (x2);
\draw[ultra thick,color=Blue] (x2) -- (x3);
\draw[ultra thick,color=Blue] (x1) -- (x4);
\draw[ultra thick,color=Blue] (x2) -- (x5);
\draw[ultra thick,color=Blue] (x3) -- (x6);
\draw[ultra thick,color=Blue] (x4) -- (x5);
\draw[ultra thick,color=Blue] (x5) -- (x6);
\draw[ultra thick,color=Blue] (x5) -- (x8);
\draw[ultra thick,color=Blue] (x6) -- (x9);
\draw[ultra thick,color=Blue] (x8) -- (x9);
\draw[ultra thick,color=Blue] (x9) -- (x12);
\draw[ultra thick,color=Blue] (x8) -- (x11);
\draw[ultra thick,color=Blue] (x11) -- (x12);
\draw[ultra thick,color=Blue] (x7) -- (x8);

%vertices:
\draw[fill,thick,color=Blue] (x11) circle (1pt);
\draw[fill,thick,color=Blue] (x1) circle (1pt);
\draw[fill,thick,color=Blue] (x2) circle (1pt);
\draw[fill,thick,color=Blue] (x3) circle (1pt);
\draw[fill,thick,color=Blue] (x4) circle (1pt);
\draw[fill,thick,color=Blue] (x5) circle (1pt);
\draw[fill,thick,color=Blue] (x6) circle (1pt);
\draw[fill,thick,color=Blue] (x7) circle (1pt);
\draw[fill,thick,color=Blue] (x8) circle (1pt);
\draw[fill,thick,color=Blue] (x9) circle (1pt);
\draw[fill,thick,color=Blue] (x12) circle (1pt);

\draw (2.3,-0.3) edge [dotted, ultra thick, color=Black, bend left =50] (3.3,0.7) node [right] {};

\end{tikzpicture}
\quad
\gamma_4:
\begin{tikzpicture}[baseline=5ex,scale=1.]
%\draw [help lines] (0,0) grid (5,4);
%coordinates:
\coordinate (x1) at (0,0) ;
\coordinate (x2) at (0,1) ;
\coordinate (x3) at (0,2) ;
\coordinate (x6) at (1,2) ;
\coordinate (x5) at (1,1) ;
\coordinate (x4) at (1,0) ;
\coordinate (x9) at (2,2) ;
\coordinate (x8) at (2,1) ;
\coordinate (x7) at (2,0) ;
\coordinate (x12) at (3,2) ;
\coordinate (x11) at (3,1) ;
\coordinate (x10) at (3,0) ;

%external momenta:
\node (p4) at (-0.5,-0.5) {$p_4$};
\node (p1) at (-0.5,2.5) {$p_1$};
\node (p3) at (3.5,-0.5) {$p_3$};
\node (p2) at (3.5,2.5) {$p_2$};

%jets:
\draw[color=green] (x1) -- (p4);
\draw[fill,thick,color=green] (x1) circle (1pt);
\draw[color=green] (x1) -- (x2);
\draw[color=green] (x1) -- (x4);
\draw[color=LimeGreen] (x3) -- (p1);
\draw[color=ForestGreen] (x12) -- (p2);
\draw[color=Green] (x10) -- (p3);

%edges:
\draw[ultra thick,color=Blue] (x2) -- (x3);
\draw[ultra thick,color=Blue] (x2) -- (x5);
\draw[ultra thick,color=Blue] (x3) -- (x6);
\draw[ultra thick,color=Blue] (x4) -- (x5);
\draw[ultra thick,color=Blue] (x5) -- (x6);
\draw[ultra thick,color=Blue] (x4) -- (x7);
\draw[ultra thick,color=Blue] (x5) -- (x8);
\draw[ultra thick,color=Blue] (x6) -- (x9);
\draw[ultra thick,color=Blue] (x7) -- (x8);
\draw[ultra thick,color=Blue] (x8) -- (x9);
\draw[ultra thick,color=Blue] (x7) -- (x10);
\draw[ultra thick,color=Blue] (x8) -- (x11);
\draw[ultra thick,color=Blue] (x9) -- (x12);
\draw[ultra thick,color=Blue] (x10) -- (x11);
\draw[ultra thick,color=Blue] (x11) -- (x12);

%vertices:
\draw[fill,thick,color=Blue] (x2) circle (1pt);
\draw[fill,thick,color=Blue] (x3) circle (1pt);
\draw[fill,thick,color=Blue] (x4) circle (1pt);
\draw[fill,thick,color=Blue] (x5) circle (1pt);
\draw[fill,thick,color=Blue] (x6) circle (1pt);
\draw[fill,thick,color=Blue] (x7) circle (1pt);
\draw[fill,thick,color=Blue] (x8) circle (1pt);
\draw[fill,thick,color=Blue] (x9) circle (1pt);
\draw[fill,thick,color=Blue] (x10) circle (1pt);
\draw[fill,thick,color=Blue] (x11) circle (1pt);
\draw[fill,thick,color=Blue] (x12) circle (1pt);

\draw (0.7,-0.3) edge [dotted, ultra thick, color=Black, bend right =50] (-0.3,0.7) node [right] {};

\end{tikzpicture}
\end{aligned}
\end{equation}
Overlaying all these subgraphs according to \emph{Step 2} leads to
\begin{equation}
\label{eq:onshell_graph_finding_result1}
\gamma_1\sqcup \gamma_2\sqcup\gamma_3\sqcup\gamma_4:\quad
\begin{tikzpicture}[baseline=5ex,scale=1.]
%\draw [help lines] (-1,-1) grid (6,4);
%coordinates:
\coordinate (x1) at (0,0) ;
\coordinate (x2) at (0,1) ;
\coordinate (x3) at (0,2) ;
\coordinate (x6) at (1,2) ;
\coordinate (x5) at (1,1) ;
\coordinate (x4) at (1,0) ;
\coordinate (x9) at (2,2) ;
\coordinate (x8) at (2,1) ;
\coordinate (x7) at (2,0) ;
\coordinate (x12) at (3,2) ;
\coordinate (x11) at (3,1) ;
\coordinate (x10) at (3,0) ;

%external momenta:
\node (p4) at (-0.5,-0.5) {$p_4$};
\node (p1) at (-0.5,2.5) {$p_1$};
\node (p3) at (3.5,-0.5) {$p_3$};
\node (p2) at (3.5,2.5) {$p_2$};

%soft
\draw[color=Red,dashed] (x6) -- (x9);
\draw[color=Red,dashed] (x5) -- (x8);
\draw[color=Red,dashed] (x10) -- (x11);
\draw[color=Red,dashed] (x1) -- (x2);

%jets:
\draw[color=green] (x1) -- (p4);
\draw[fill,thick,color=green] (x1) circle (1pt);
\draw[green,green] (x1) -- (x4);
\draw[color=LimeGreen] (x3) -- (p1);
\draw[color=LimeGreen] (x3) -- (x6);
\draw[color=LimeGreen] (x2) -- (x3);

\draw[color=LimeGreen] (x2) -- (x5);
\draw[color=LimeGreen] (x5) -- (x6);
\draw[color=LimeGreen] (x4) -- (x5);
\draw[fill,thick,color=LimeGreen] (x6) circle (1pt);
\draw[fill,thick,color=LimeGreen] (x2) circle (1pt);
\draw[fill,thick,color=LimeGreen] (x3) circle (1pt);
\draw[fill,thick,color=LimeGreen] (x5) circle (1pt);

\draw[color=ForestGreen] (x12) -- (p2);
\draw[fill,thick,color=ForestGreen] (x8) circle (1pt);
\draw[fill,thick,color=ForestGreen] (x9) circle (1pt);
\draw[fill,thick,color=ForestGreen] (x11) circle (1pt);
\draw[fill,thick,color=ForestGreen] (x12) circle (1pt);
\draw[color=ForestGreen] (x7) -- (x8);
\draw[color=ForestGreen] (x8) -- (x9);
\draw[color=ForestGreen] (x8) -- (x11);
\draw[color=ForestGreen] (x9) -- (x12);

\draw[color=ForestGreen] (x11) -- (x12);

\draw[color=Green] (x10) -- (p3);
\draw[fill,thick,color=Green] (x10) circle (1pt);
\draw[color=Green] (x7) -- (x10);

%edges:
\draw[ultra thick,color=Blue] (x4) -- (x7);

%vertices:
\draw[fill,thick,color=Blue] (x4) circle (1pt);
\draw[fill,thick,color=Blue] (x7) circle (1pt);
\end{tikzpicture},
\end{equation}
where the soft subgraph consists of four disconnected components. It is straightforward to verify that the hard and jet subgraphs are all connected, and $H\cup J\setminus J_i$ is mojetic for each $i\in \{1,2,3,4\}$. As a result, the conditions of Step 3 are automatically satisfied, implying that the configuration in eq.~(\ref{eq:onshell_graph_finding_result1}) is a valid region. By considering all the possible choices of $\{\gamma_1,\gamma_2,\gamma_3,\gamma_4\}$, one obtains the complete list of regions that appear in the on-shell expansion.

We emphasize that the output of \emph{Step 1} and \emph{2} should be discarded if it does not satisfy any of the conditions in \emph{Step 3}. Examples of such cases are given in section 4.3 of ref.~\cite{Gardi:2022khw}.

\subsection{Regions for the single-soft expansion with two colored partons}
\label{section-regions_1_234_decay}

We now consider the soft expansion for the decay processes of eq.~(\ref{eq:p1_to_p2p3p4_decay}), where $p_1$ is off-shell while $p_2$, $p_3$, and $p_4$ are on shell. Moreover, $p_2$ and $p_3$ are in distinct directions and $p_4$ is soft. The kinematic limit can be summarized as
\begin{subequations}
\label{eq:Sudakov_soft_gluon_kinematics}
    \begin{align}
        & p_1^2\sim Q^2,\qquad p_2^2=p_3^2=p_4^2=0,\\
        & p_1\cdot p_4\sim p_2\cdot p_4\sim p_3\cdot p_4 \sim \lambda Q^2,\qquad p_1\cdot p_2\sim Q^2,
    \end{align}
\end{subequations}
which is a special case of soft expansion described in the beginning of this section in eq.~(\ref{eq:wideangle_soft_gluon_kinematics}). 

Note that in this particular case, where $p_4$ is the unique soft external momentum, additional requirements of the configurations of $H$, $J$, and $S$ are needed~\cite{Ma23}, giving the following possibilities.
\begin{itemize}
    \item If there are no internal soft propagators, then there can be at most one nontrivial\footnote{A jet is called nontrivial if it has one or more edges.} jet $J_i$ ($i=2$ or $3$), to which $p_4$ directly attached. In the special case that neither $J_2$ or $J_3$ is nontrivial, the region is referred to as the \emph{hard region}, where all the loop momenta are equally off shell.
        
    \item If there are internal soft propagators, then each component of $S$ must be adjacent to both $J_2$ and $J_3$. In addition, $p_4$ must enter a soft vertex.
    
    In general, such regions are depicted in figure~\ref{soft_Sudakov_classical_picture}, where the hard, jet, and soft subgraphs satisfy:
    \begin{enumerate}
        \item [a)] the hard subgraph $H$ is connected and attached by $p_1$;
        \item [b)] the jet subgraphs $J_2$ and $J_3$ are both connected and adjacent to $H$, and are attached by $p_2$ and $p_3$ respectively;
        \item [c)] the soft subgraph $S$ is attached by $p_4$, and each of its connected components is adjacent to both $J_2$ and $J_3$.
    \end{enumerate}
    \begin{figure}[t]
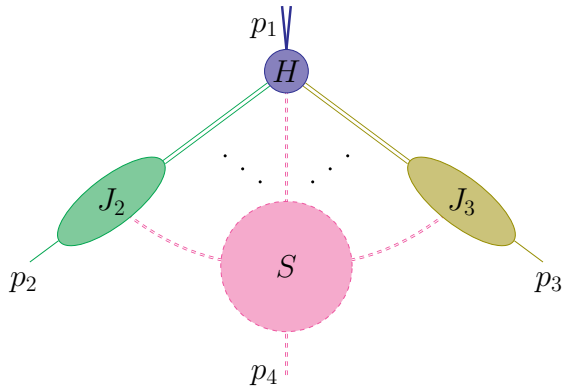

    \centering
    \include{figs/soft_Sudakov_classical_picture}
    \vspace{1em}
    \caption{The general configuration of regions in the process of eq.~(\ref{eq:p1_to_p2p3p4_decay}), where there are internal soft propagators. The external momenta $p_1$, $p_2$, $p_3$, and $p_4$ attach to $H$, $J_2$, $J_3$, and $S$, respectively.}
    \label{soft_Sudakov_classical_picture}
    \end{figure}
\end{itemize}
This is illustrated below with some examples of regions (marked $\greencheckmark[ForestGreen]$) and non-region configurations (marked $\crossmark[Red]$).
\begin{equation}
\begin{aligned}
\begin{tikzpicture}[baseline=11ex, scale=0.4]
% \draw [help lines] (0,0) grid (10,10);
\path (2,7) edge [Green, thick] (5,3) {};
\path (5,6.5) edge [Red, thick] (5,5) {};
\path (2.75,6) edge [Green, thick, bend left = 10] (6,4.33) {};
\path (3.5,5) edge [Green, thick, bend left = 10] (5.5,3.67) {};
\path (8,7) edge [LimeGreen, thick] (6,4.33) {};
\path (6,4.33) edge [Blue, ultra thick] (5,3) {};
\path (4.8,1.5) edge [Blue, ultra thick] (5,3) {};
\path (5.2,1.5) edge [Blue, ultra thick] (5,3) {};
\node [draw, circle, minimum size=3pt, color=Green, fill=Green, inner sep=0pt, outer sep=0pt] () at (2.75,6) {};
\node [draw, circle, minimum size=3pt, color=Blue, fill=Blue, inner sep=0pt, outer sep=0pt] () at (5.5,3.67) {};
\node [draw, circle, minimum size=3pt, color=Green, fill=Green, inner sep=0pt, outer sep=0pt] () at (3.5,5) {};
\node [draw, circle, minimum size=3pt, color=Blue, fill=Blue, inner sep=0pt, outer sep=0pt] () at (6,4.33) {};
\node [draw, circle, minimum size=3pt, color=Green, fill=Green, inner sep=0pt, outer sep=0pt] () at (5,5) {};
\node [draw, circle, minimum size=3pt, color=Blue, fill=Blue, inner sep=0pt, outer sep=0pt] () at (5,3) {};
\node () at (4.4,2.2) {\color{Blue} $q_1$};
\node () at (5,7) {\color{Red} $l_1$};
\node () at (8,7.5) {\color{LimeGreen} $p_2$};
\node () at (2,7.5) {\color{Green} $p_1$};
\node () at (5,0.5) {\large (a)};
\node () at (6.4,2.4) {$\greencheckmark[ForestGreen]$};
\end{tikzpicture}\quad\ \ 
\begin{tikzpicture}[baseline=11ex, scale=0.4]
% \draw [help lines] (0,0) grid (10,10);
\path (2,7) edge [Green, thick] (5,3) {};
\path (5,7) edge [Red, thick, bend left = 0] (5,8.1) {};
\path (2.75,6) edge [Red, thick, bend left = 50] (7.25,6) {};
\path (3.5,5) edge [Green, thick, bend left = 10] (5.33,3.5) {};
\path (8,7) edge [LimeGreen, thick] (5,3) {};
\path (4.8,1.5) edge [Blue, ultra thick] (5,3) {};
\path (5.2,1.5) edge [Blue, ultra thick] (5,3) {};
\path (5.33,3.5) edge [Blue, ultra thick] (5,3) {};
\node [draw, circle, minimum size=3pt, color=Green, fill=Green, inner sep=0pt, outer sep=0pt] () at (2.75,6) {};
\node [draw, circle, minimum size=3pt, color=LimeGreen, fill=LimeGreen, inner sep=0pt, outer sep=0pt] () at (7.25,6) {};
\node [draw, circle, minimum size=3pt, color=Green, fill=Green, inner sep=0pt, outer sep=0pt] () at (3.5,5) {};
\node [draw, circle, minimum size=3pt, color=Blue, fill=Blue, inner sep=0pt, outer sep=0pt] () at (5.33,3.5) {};
\node [draw, circle, minimum size=3pt, color=Blue, fill=Blue, inner sep=0pt, outer sep=0pt] () at (5,3) {};
\node [draw, circle, minimum size=3pt, color=Red, fill=Red, inner sep=0pt, outer sep=0pt] () at (5,7) {};
\node () at (4.4,2.2) {\color{Blue} $q_1$};
\node () at (5.5,8) {\color{Red} $l_1$};
\node () at (8,7.5) {\color{LimeGreen} $p_2$};
\node () at (2,7.5) {\color{Green} $p_1$};
\node () at (5,0.5) {\large (b)};
\node () at (6.4,2.4) {$\greencheckmark[Green]$};
\end{tikzpicture}\quad
\begin{tikzpicture}[baseline=11ex, scale=0.4]
% \draw [help lines] (0,0) grid (10,10);
\path (2,7) edge [Green, thick] (5,3) {};
\path (5,7.7) edge [Red, thick] (5,6.25) {};
\path (2.75,6) edge [Red, thick, bend left = 10] (7.25,6) {};
\path (3.5,5) edge [Red, thick, bend left = 10] (6.5,5) {};
\path (8,7) edge [LimeGreen, thick] (5,3) {};
\path (4.8,1.5) edge [Blue, ultra thick] (5,3) {};
\path (5.2,1.5) edge [Blue, ultra thick] (5,3) {};
\node [draw, circle, minimum size=3pt, color=Green, fill=Green, inner sep=0pt, outer sep=0pt] () at (2.75,6) {};
\node [draw, circle, minimum size=3pt, color=LimeGreen, fill=LimeGreen, inner sep=0pt, outer sep=0pt] () at (7.25,6) {};
\node [draw, circle, minimum size=3pt, color=Green, fill=Green, inner sep=0pt, outer sep=0pt] () at (3.5,5) {};
\node [draw, circle, minimum size=3pt, color=LimeGreen, fill=LimeGreen, inner sep=0pt, outer sep=0pt] () at (6.5,5) {};
\node [draw, circle, minimum size=3pt, color=Red, fill=Red, inner sep=0pt, outer sep=0pt] () at (5,6.25) {};
\node [draw, circle, minimum size=3pt, color=Blue, fill=Blue, inner sep=0pt, outer sep=0pt] () at (5,3) {};
\node () at (4.4,2.2) {\color{Blue} $q_1$};
\node () at (5,8.2) {\color{Red} $l_1$};
\node () at (8,7.5) {\color{LimeGreen} $p_2$};
\node () at (2,7.5) {\color{Green} $p_1$};
\node () at (5,0.5) {\large (c)};
\node () at (6.4,2.4) {$\greencheckmark[ForestGreen]$};
\end{tikzpicture}\quad
\begin{tikzpicture}[baseline=11ex, scale=0.4]
% \draw [help lines] (0,0) grid (10,10);
\path (2,7) edge [Green, thick] (5,3) {};
\path (5,7) edge [Red, thick] (5,5.5) {};
\path (2.75,6) edge [Red, thick, bend left = 20] (5,5.5) {};
\path (4,4.33) edge [Red, thick, bend right = 20] (5,5.5) {};
\path (7.25,6) edge [Red, thick, bend right = 20] (5,5.5) {};
\path (8,7) edge [LimeGreen, thick] (5,3) {};
\path (4.8,1.5) edge [Blue, ultra thick] (5,3) {};
\path (5.2,1.5) edge [Blue, ultra thick] (5,3) {};
\node [draw, circle, minimum size=3pt, color=Green, fill=Green, inner sep=0pt, outer sep=0pt] () at (2.75,6) {};
\node [draw, circle, minimum size=3pt, color=LimeGreen, fill=LimeGreen, inner sep=0pt, outer sep=0pt] () at (7.25,6) {};
\node [draw, circle, minimum size=3pt, color=Green, fill=Green, inner sep=0pt, outer sep=0pt] () at (4,4.33) {};
\node [draw, circle, minimum size=3pt, color=Red, fill=Red, inner sep=0pt, outer sep=0pt] () at (5,5.5) {};
\node [draw, circle, minimum size=3pt, color=Blue, fill=Blue, inner sep=0pt, outer sep=0pt] () at (5,3) {};
\node () at (4.4,2.2) {\color{Blue} $q_1$};
\node () at (5,7.5) {\color{Red} $l_1$};
\node () at (8,7.5) {\color{LimeGreen} $p_2$};
\node () at (2,7.5) {\color{Green} $p_1$};
\node () at (5,0.5) {\large (d)};
\node () at (6.4,2.4) {$\greencheckmark[ForestGreen]$};
\end{tikzpicture}\\
\begin{tikzpicture}[baseline=11ex, scale=0.4]
% \draw [help lines] (0,0) grid (10,10);
\path (2,7) edge [Green, thick] (5,3) {};
\path (2.75,6) edge [Green, thick, bend left = 10] (5,4) {};
\path (4.5,7) edge [Red, thick, bend left = 10] (4.1,5) {};
\path (5,4) edge [LimeGreen, thick, bend left = 10] (7.25,6) {};
\path (8,7) edge [LimeGreen, thick] (5,3) {};
\path (5.33,3.5) edge [Blue, ultra thick] (5,4) {};
\path (5.33,3.5) edge [Blue, ultra thick] (5,3) {};
\path (4.8,1.5) edge [Blue, ultra thick] (5,3) {};
\path (5.2,1.5) edge [Blue, ultra thick] (5,3) {};
\node [draw, circle, minimum size=3pt, color=Green, fill=Green, inner sep=0pt, outer sep=0pt] () at (2.75,6) {};
\node [draw, circle, minimum size=3pt, color=LimeGreen, fill=LimeGreen, inner sep=0pt, outer sep=0pt] () at (7.25,6) {};
\node [draw, circle, minimum size=3pt, color=Green, fill=Green, inner sep=0pt, outer sep=0pt] () at (4.1,5) {};
\node [draw, circle, minimum size=3pt, color=Blue, fill=Blue, inner sep=0pt, outer sep=0pt] () at (5,3) {};
\node [draw, circle, minimum size=3pt, color=Blue, fill=Blue, inner sep=0pt, outer sep=0pt] () at (5,4) {};
\node [draw, circle, minimum size=3pt, color=Blue, fill=Blue, inner sep=0pt, outer sep=0pt] () at (5.33,3.5) {};
\node () at (4.4,2.2) {\color{Blue} $q_1$};
\node () at (4,7.5) {\color{Red} $l_1$};
\node () at (8.5,7.5) {\color{LimeGreen} $p_2$};
\node () at (1.5,7.5) {\color{Green} $p_1$};
\node () at (5,0.5) {\large (a')};
\node () at (6.4,2.4) {$\crossmark[Red]$};
\end{tikzpicture}\quad
\begin{tikzpicture}[baseline=11ex, scale=0.4]
% \draw [help lines] (0,0) grid (10,10);
\path (2,7) edge [Green, thick] (5,3) {};
\path (5,6.2) edge [Red, thick, bend left = 20] (4.5,4.3) {};
\path (2.75,6) edge [Red, thick, bend left = 50] (7.25,6) {};
\path (3.5,5) edge [Green, thick, bend left = 10] (5.33,3.5) {};
\path (8,7) edge [LimeGreen, thick] (5,3) {};
\path (4.8,1.5) edge [Blue, ultra thick] (5,3) {};
\path (5.2,1.5) edge [Blue, ultra thick] (5,3) {};
\path (5.33,3.5) edge [Blue, ultra thick] (5,3) {};
\node [draw, circle, minimum size=3pt, color=Green, fill=Green, inner sep=0pt, outer sep=0pt] () at (2.75,6) {};
\node [draw, circle, minimum size=3pt, color=LimeGreen, fill=LimeGreen, inner sep=0pt, outer sep=0pt] () at (7.25,6) {};
\node [draw, circle, minimum size=3pt, color=Green, fill=Green, inner sep=0pt, outer sep=0pt] () at (3.5,5) {};
\node [draw, circle, minimum size=3pt, color=Blue, fill=Blue, inner sep=0pt, outer sep=0pt] () at (5.33,3.5) {};
\node [draw, circle, minimum size=3pt, color=Blue, fill=Blue, inner sep=0pt, outer sep=0pt] () at (5,3) {};
\node [draw, circle, minimum size=3pt, color=Green, fill=Green, inner sep=0pt, outer sep=0pt] () at (4.5,4.3) {};
\node () at (4.4,2.2) {\color{Blue} $q_1$};
\node () at (4.6,5.9) {\color{Red} $l_1$};
\node () at (8,7.5) {\color{LimeGreen} $p_2$};
\node () at (2,7.5) {\color{Green} $p_1$};
\node () at (5,0.5) {\large (b')};
\node () at (6.4,2.4) {$\crossmark[Red]$};
\end{tikzpicture}\quad
\begin{tikzpicture}[baseline=11ex, scale=0.4]
% \draw [help lines] (0,0) grid (10,10);
\path (2,7) edge [Green, thick] (5,3) {};
\path (2.75,6) edge [Red, thick, bend left = 30] (7.25,6) {};
\path (3.5,5) edge [Red, thick, bend left = 10] (6.5,5) {};
\path (2.75,6) edge [Red, thick] (3.5,7) {};
\path (8,7) edge [LimeGreen, thick] (5,3) {};
\path (4.8,1.5) edge [Blue, ultra thick] (5,3) {};
\path (5.2,1.5) edge [Blue, ultra thick] (5,3) {};
\node [draw, circle, minimum size=3pt, color=Green, fill=Green, inner sep=0pt, outer sep=0pt] () at (2.75,6) {};
\node [draw, circle, minimum size=3pt, color=LimeGreen, fill=LimeGreen, inner sep=0pt, outer sep=0pt] () at (7.25,6) {};
\node [draw, circle, minimum size=3pt, color=Green, fill=Green, inner sep=0pt, outer sep=0pt] () at (3.5,5) {};
\node [draw, circle, minimum size=3pt, color=LimeGreen, fill=LimeGreen, inner sep=0pt, outer sep=0pt] () at (6.5,5) {};
\node [draw, circle, minimum size=3pt, color=Blue, fill=Blue, inner sep=0pt, outer sep=0pt] () at (5,3) {};
\node () at (4.4,2.2) {\color{Blue} $q_1$};
\node () at (4,7.5) {\color{Red} $l_1$};
\node () at (8,7.5) {\color{LimeGreen} $p_2$};
\node () at (2,7.5) {\color{Green} $p_1$};
\node () at (5,0.5) {\large (c')};
\node () at (6.4,2.4) {$\crossmark[Red]$};
\end{tikzpicture}\quad
\begin{tikzpicture}[baseline=11ex, scale=0.4]
% \draw [help lines] (0,0) grid (10,10);
\path (2,7) edge [Green, thick] (5,3) {};
\path (2.75,6) edge [Red, thick] (3.5,7) {};
\path (2.75,6) edge [Red, thick, bend left = 20] (5,5.5) {};
\path (4,4.33) edge [Red, thick, bend right = 20] (5,5.5) {};
\path (7.25,6) edge [Red, thick, bend right = 20] (5,5.5) {};
\path (8,7) edge [LimeGreen, thick] (5,3) {};
\path (4.8,1.5) edge [Blue, ultra thick] (5,3) {};
\path (5.2,1.5) edge [Blue, ultra thick] (5,3) {};
\node [draw, circle, minimum size=3pt, color=Green, fill=Green, inner sep=0pt, outer sep=0pt] () at (2.75,6) {};
\node [draw, circle, minimum size=3pt, color=LimeGreen, fill=LimeGreen, inner sep=0pt, outer sep=0pt] () at (7.25,6) {};
\node [draw, circle, minimum size=3pt, color=Green, fill=Green, inner sep=0pt, outer sep=0pt] () at (4,4.33) {};
\node [draw, circle, minimum size=3pt, color=Red, fill=Red, inner sep=0pt, outer sep=0pt] () at (5,5.5) {};
\node [draw, circle, minimum size=3pt, color=Blue, fill=Blue, inner sep=0pt, outer sep=0pt] () at (5,3) {};
\node () at (4.4,2.2) {\color{Blue} $q_1$};
\node () at (3.5,7.5) {\color{Red} $l_1$};
\node () at (8,7.5) {\color{LimeGreen} $p_2$};
\node () at (2,7.5) {\color{Green} $p_1$};
\node () at (5,0.5) {\large (d')};
\node () at (6.4,2.4) {$\crossmark[Red]$};
\end{tikzpicture}
\end{aligned}
\label{eq:threshold_motivation_Sudakov_collinear_regions}
\end{equation}
In particular, the \emph{maximally soft region} feature the following configuration: there is a unique path connecting $p_1$ and $p_2$ and carrying momenta that are either hard or collinear to $p_2$; meanwhile there is a unique path connecting $p_1$ and $p_2$ and carrying momenta that are either hard or collinear to $p_3$; all the remaining propagators carry soft momenta. Examples of maximally soft regions include graphs (c) and (d) in (\ref{eq:threshold_motivation_Sudakov_collinear_regions}).

To expand a particular Feynman integral, we first find all the allowed maximally soft regions. For each of these regions we then need to find a loop momentum parameterization, such that the leading term of the propagators associated with the edges of the corresponding graph in that region scale accordingly. Loop momentum parameterizations doing this job are not unique, but can be generated straightforwardly, by checking that the spanning tree $T$ dual to the edges carrying the chosen loop momenta has the \textit{maximum weight} $\omega_T$, according to the definition:
\begin{equation}
\omega_T(R)\equiv\sum_{e\notin{T}} \omega_e(R)\,.
\end{equation}
Here
$\omega_e=0,1,2$ if the edge $e$ belongs to a hard, jet or soft subgraph respectively. A simple and relatively cheap algorithm is given by just computing $\omega_T$ for all the possible spanning trees and keeping only those loop momentum parameterizations with the maximum weight. Any one of these will suffice.

Performing an explicit scaling of all the soft loop momenta with the expansion parameter $\lambda$ then allows us to perform a simple Laurent series expansion at the integrand level. Finally, we may set $\lambda=1$ and perform the integration over loop momenta, as discussed in the previous section. 
We would like to emphasize that our method allows us to expand beyond the leading power in the soft expansion. 
In general, it allows us to compute the expansion of our scattering amplitudes to as high a power as desired.

Our analyses above is sufficient to develop soft expansions for the $p_1\to p_2p_3p_4$ process, namely, eq.~(\ref{eq:Sudakov_soft_gluon_kinematics}). Based on the findings of ref.~\cite{Ma23}, our method can be readily extended to soft expansions for generic wide-angle scattering, eq.~(\ref{eq:wideangle_soft_gluon_kinematics}).

%% file: figs/soft_classical_picture.tex
\resizebox{\textwidth}{!}{
\begin{tikzpicture}[baseline=11ex, line width = 0.6, font=\huge, mydot/.style={circle, fill, inner sep=.7pt}]
% \draw [help lines] (0,0) grid (10,10);

\path (6,8) edge [double,double distance=2pt,color=Green] (2,5) {};
\path (6,8) edge [double,double distance=2pt,color=LimeGreen] (6,3.5) {};
\path (6,8) edge [double,double distance=2pt,color=olive] (10,5) {};

\draw (3,2) edge [dashed,double,color=Rhodamine] (6,8) node [right] {};
\draw (3,2) edge [dashed,double,color=Rhodamine,bend left = 15] (2,5) {};
\draw (3,2) edge [dashed,double,color=Rhodamine,bend right = 15] (6,3.5) {};
\draw (3,2) edge [dashed,double,color=Rhodamine,bend right = 40] (10,5) {};

\draw (3,8.5) edge [dashed,double,color=Rhodamine,bend left = 15] (5.5,8) {};
\draw (0,7) edge [dashed,double,color=Rhodamine,bend left = 15] (2,5.6) {};
\draw (1,1.5) edge [dashed,double,color=Rhodamine,bend left = 30] (3,2) {};
\draw (2.5,0) edge [dashed,double,color=Rhodamine,bend right = 30] (3,2) {};

\node (q1) at (4.5,9.5) {};
\node (q1p) at (4.7,9.7) {};
\node (qn) at (7.5,9.5) {};
\node (qnp) at (7.3,9.7) {};
\draw (q1) edge [color=Blue] (6,8) node [] {};
\draw (q1p) edge [color=Blue] (6,8) node [left] {$q_1$};
\draw (qn) edge [color=Blue] (6,8) node [] {};
\draw (qnp) edge [color=Blue] (6,8) node [right] {$q_L$};

\node (p1) at (0,3.5) {};
\node (p2) at (6,1) {};
\node (pn) at (12,3.5) {};
\draw (p1) edge [color=Green] (2,5) node [below] {$p_1$};
\draw (p2) edge [color=LimeGreen] (6,3.5) node [below] {$p_2$};
\draw (pn) edge [color=olive] (10,5) node [below] {$p_K$};

\path (6,3.5)-- node[mydot, pos=.333] {} node[mydot] {} node[mydot, pos=.666] {}(10,5);
\path (q1)-- node[mydot, pos=.333] {} node[mydot] {} node[mydot, pos=.666] {}(qn);
\path (1,6.5)-- node[mydot, pos=.333] {} node[mydot] {} node[mydot, pos=.666] {}(4,8.5);
\path (1,2)-- node[mydot, pos=.3] {} node[mydot] {} node[mydot, pos=.7] {}(3,0);

\node[draw=Blue,circle,minimum size=1cm,fill=Blue!50] () at (6,8){};
\node[dashed, draw=Rhodamine,circle,minimum size=1.8cm,fill=Rhodamine!50] () at (3,2){};
\node[draw=Green,ellipse,minimum height=3cm, minimum width=1.1cm,fill=Green!50,rotate=-52] () at (2,5){};
\node[draw=LimeGreen,ellipse,minimum height=3cm, minimum width=1.1cm,fill=LimeGreen!50] () at (6,3.5){};
\node[draw=olive,ellipse,minimum height=3cm, minimum width=1.1cm,fill=olive!50,rotate=52] () at (10,5){};

\node at (6,8) {$H$};
\node at (3,2) {$S$};
\node at (2,5) {$J_1$};
\node at (6,3.5) {$J_2$};
\node at (10,5) {$J_K$};
\node at (2.5,8.5) {$l_1$};
\node at (1.9,0) {$l_M$};

\end{tikzpicture}
}
\vspace{-2em}

%% file: figs/onshell_classical_picture.tex
\resizebox{\textwidth}{!}{
\begin{tikzpicture}[baseline=11ex, line width = 0.6, font=\huge, mydot/.style={circle, fill, inner sep=.7pt}]
%\draw [help lines] (-1,1) grid (12,10);

\node[draw=Blue,circle,minimum size=1cm,fill=Blue!50] (h) at (6,8){};
\node[dashed, draw=Rhodamine,circle,minimum size=1.8cm,fill=Rhodamine!50] (s) at (3,2){};
\node[draw=Green,ellipse,minimum height=3cm, minimum width=1.1cm,fill=Green!50,rotate=-52] (j1) at (2,5){};
\node[draw=LimeGreen,ellipse,minimum height=3cm, minimum width=1.1cm,fill=LimeGreen!50] (j2) at (6,3.5){};
\node[draw=olive,ellipse,minimum height=3cm, minimum width=1.1cm,fill=olive!50,rotate=52] (jn) at (10,5){};

\node at (h) {$H$};
\node at (s) {$S$};
\node at (j1) {$J_1$};
\node at (j2) {$J_2$};
\node at (jn) {$J_K$};

\path (h) edge [double,double distance=2pt,color=Green] (j1) {};
\path (h) edge [double,double distance=2pt,color=LimeGreen] (j2) {};
\path (h) edge [double,double distance=2pt,color=olive] (jn) {};

\draw (s) edge [dashed,double,color=Rhodamine] (h) node [right] {};
\draw (s) edge [dashed,double,color=Rhodamine,bend left = 15] (j1) {};
\draw (s) edge [dashed,double,color=Rhodamine,bend right = 15] (j2) {};
\draw (s) edge [dashed,double,color=Rhodamine,bend right = 40] (jn) {};

\node (q1) at (4.5,9.5) {};
\node (q1p) at (4.7,9.7) {};
\node (qn) at (7.5,9.5) {};
\node (qnp) at (7.3,9.7) {};
\draw (q1) edge [color=Blue] (h) node [] {};
\draw (q1p) edge [color=Blue] (h) node [left] {$q_1$};
\draw (qn) edge [color=Blue] (h) node [] {};
\draw (qnp) edge [color=Blue] (h) node [right] {$q_L$};

\node (p1) at (0,3.5) {};
\node (p2) at (6,1) {};
\node (pn) at (12,3.5) {};
\draw (p1) edge [color=Green] (j1) node [below] {$p_1$};
\draw (p2) edge [color=LimeGreen] (j2) node [below] {$p_2$};
\draw (pn) edge [color=olive] (jn) node [below] {$p_K$};

\path (j2)-- node[mydot, pos=.333] {} node[mydot] {} node[mydot, pos=.666] {}(jn);
\path (q1)-- node[mydot, pos=.333] {} node[mydot] {} node[mydot, pos=.666] {}(qn);

\end{tikzpicture}
}
\vspace{-2em}

%% file: figs/soft_Sudakov_classical_picture.tex
\resizebox{0.5\textwidth}{!}{
\begin{tikzpicture}[baseline=11ex, line width = 0.6, font=\huge, mydot/.style={circle, fill, inner sep=.7pt}]
% \draw [help lines] (0,0) grid (10,10);

\path (h) edge [double,double distance=2pt,color=Green] (j1) {};
\path (h) edge [double,double distance=2pt,color=olive] (jn) {};

\draw (6,3.5) edge [dashed, double, color=Rhodamine, bend left = 20] (2,5) {};
\draw (6,3.5) edge [dashed, double, color=Rhodamine, bend right = 20] (10,5) {};
\draw (6,1) edge [dashed, double, color=Rhodamine] (6,3.5) {};
\draw (6,8) edge [dashed, double, color=Rhodamine] (6,3.5) {};

\node[draw=Blue,circle,minimum size=1cm,fill=Blue!50] (h) at (6,8){};
\node[dashed, draw=Rhodamine,circle,minimum size=3cm,fill=Rhodamine!50] (s) at (6,3.5){};
\node[draw=Green,ellipse,minimum height=3cm, minimum width=1.1cm,fill=Green!50,rotate=-52] (j1) at (2,5){};
\node[draw=olive,ellipse,minimum height=3cm, minimum width=1.1cm,fill=olive!50,rotate=52] (jn) at (10,5){};

\node at (h) {$H$};
\node at (s) {$S$};
\node at (j1) {$J_2$};
\node at (jn) {$J_3$};
\node at (5.5,9) {$p_1$};
\node at (5.5,1) {$p_4$};

\node (q1) at (4.5,9.5) {};
\node (q1p) at (4.7,9.7) {};
\node (qn) at (7.5,9.5) {};
\node (qnp) at (7.3,9.7) {};
\draw (5.9,9.5) edge [ultra thick, color=Blue] (6,8.5) node [] {};
\draw (6.1,9.5) edge [ultra thick, color=Blue] (6,8.5) node [] {};

\node (p1) at (0,3.5) {};
\node (p2) at (6,1) {};
\node (pn) at (12,3.5) {};
\draw (p1) edge [color=Green] (j1) node [below] {$p_2$};
\draw (pn) edge [color=olive] (jn) node [below] {$p_3$};

\path (4,6.5)-- node[mydot, pos=.3] {} node[mydot] {} node[mydot, pos=.7] {}(6,5);
\path (8,6.5)-- node[mydot, pos=.3] {} node[mydot] {} node[mydot, pos=.7] {}(6,5);

\end{tikzpicture}
}
\vspace{-2em}

%% file: Chapters/IRSubtraction.tex
\section{Renormalization and infrared pole structure}
\label{sec:IR}

In this section, we briefly describe the renormalization and subtraction of singularities of the bare emission current.
In general, the infrared and ultraviolet singularities of a scattering amplitude computed in perturbative QCD can be subtracted to yield a finite remainder using the following definitions.
\beq
\label{eq:subtdiv}
\mathcal{A}_f(\alpha_S(\mu^2),\{p_i\})={\mathbf Z}(\alpha_S(\mu^2),\{p_i\},\epsilon)\ {\mathbf Z}_{\text{UV}}\ \mathcal{A}(\{p_i\},\epsilon).
\eeq
The factor ${\mathbf Z}_{\text{UV}}$ implements the renormalization of the strong coupling constant in the $\overline{\text{MS}}$ scheme and $\epsilon$ is the dimensional regulator related to the space-time dimension by $d=4-2\epsilon$.
\beq
{\mathbf Z}_{\text{UV}}\alpha_S=\alpha_S(\mu^2) \left(\frac{\mu^2}{4\pi}\right)^{-\epsilon} e^{\gamma_E \epsilon} Z_{\alpha_S}.
\eeq
The factor $Z_{\alpha_S}$ is given in terms of the $\beta$-function~\cite{Baikov:2016tgj,Czakon:2004bu,Herzog:2017ohr,Larin:1993tp,Tarasov:1980au,vanRitbergen:1997va} through three loops in QCD by the following expression.
\bea
Z_{\alpha_S}&=&1
-\frac{\alpha_S(\mu^2)}{\pi} \frac{1}{\epsilon}\beta_0
+\left(\frac{\alpha_S(\mu^2)}{\pi} \right)^2\left( \frac{1}{\epsilon^2}\beta_0^2-\frac{1}{2\epsilon} \beta_1\right)\\
&&-\left(\frac{\alpha_S(\mu^2)}{\pi} \right)^3\left( \frac{1}{\epsilon^3}\beta_0^3-\frac{7}{6\epsilon^2}\beta_0 \beta_1+\frac{1}{3\epsilon}\beta_2 \right)+\mathcal{O}(\alpha_S^4).\nonumber
\eea
The factor ${\mathbf Z}(\alpha_S(\mu^2),\{p_i\},\epsilon) $ is an operator in color space and implements the universal subtraction of infrared and collinear singularities of loop amplitudes~\cite{Almelid:2015jia,Aybat:2006mz,Aybat:2006wq,Catani:1998bh,Dixon:2008gr,Korchemsky:1987wg,Sterman:2002qn,Becher:2019avh}.
It can be expressed in terms of the \emph{soft anomalous dimension matrix} ${\mathbf \Gamma(\alpha_S(\mu^2),\{p_i \}, \epsilon)}$ by the following path ordered exponential.
\beq
\label{eq:Zexp}
{\mathbf Z}(\alpha_S(\mu^2),\{p_i\},\epsilon) = \mathcal{P} \exp\left\{-\frac{1}{4}\int_{0}^{\mu^2} \frac{d\mu^{\prime 2}}{\mu^{\prime 2}} {\mathbf \Gamma(\alpha_S(\mu^{\prime 2}),\{p_i \}, \epsilon)} \right\}\,,
\eeq
with
\beq
\label{eq:GammaSoftDef}
 {\mathbf \Gamma(\alpha_S(\mu^2),\{p_i \}, \epsilon)} = \sum_{i\neq j} {\mathbf T}_i^a {\mathbf T}_j^a \Gamma_{\text{cusp}}(\alpha_S(\mu^2)) \log \frac{-s_{ij}}{\mu^2}+ \frac{1}{2}\sum_i \mathbf{\mathds{1}} \gamma^{R_i}_c+\bold{\Delta}(\alpha_S(\mu^2),\{p_i \}).
\eeq
Above, $\Gamma_\text{cusp}$ refers to the cusp anomalous dimension~\cite{Korchemsky:1987wg}, which is currently known exactly through four-loop order~\cite{Henn:2019swt,vonManteuffel:2020vjv}, and approximately at five loops \cite{Herzog:2018kwj}. Furthermore, $\gamma_c^R$ is the collinear anomalous dimension, obtained through four-loop order in refs.~\cite{Agarwal:2021zft,vonManteuffel:2020vjv}. The formula above was derived and calculated through three-loop order in ref.~\cite{Almelid:2015jia} and verified in $\mathcal{N}=4$ super Yang-Mills theory~\cite{Henn:2016jdu} and QCD~\cite{Caola:2022dfa,Caola:2021izf,Caola:2021rqz}. In ref.~\cite{Becher:2019avh}, its general structure was determined at four-loop order.
The term $\bold{\Delta}(\alpha_S(\mu^2),\{p_i \})$ is known as the correction to the dipole formula and starts at three-loop order. As the name suggests, it contains contributions where the color operator acts on more than two color-charged external particles simultaneously for the first time. 
This term can be further decomposed as follows.
\beq
\mathbf{\Delta}(\alpha_S(\mu^2),\{p_i \})=\left(\frac{\alpha_S(\mu^2)}{\pi}\right)^3 \left[ \mathbf{\Delta}^{(3)}_3+\mathbf{\Delta}^{(3)}_4(\{p_i\}) \right]+\mathcal{O}\left(\alpha_S^4\right).
\eeq
The expression $\mathbf{\Delta}^{(3)}_4(\{p_i\}) $ is known as the quadruple correction and involves color correlations among four or more different color-charged particles. 
Consequently, this term will not contribute to the scattering amplitudes we consider here. 
The term $\mathbf{\Delta}^{(3)}_3$ relates three different color-charged external particles and is explicitly given by
\beq
\bold{\Delta}_3^{(3)}=  \frac{1}{4} C \,f_{abe}f_{cde}  \sum_{i\neq j,i\neq k,j\neq k}\left\{{\rm \bf T}_i^a,  {\rm \bf T}_i^d\right\}   {\rm \bf T}_j^b {\rm \bf T}_k^c \,,
\eeq
with the constant $C=\zeta_5+2\zeta_3\zeta_2$.
The color operators $\bold{T}_i^a$ are defined below, via their actions on an outgoing quark, antiquark and gluon.
\bea
\bold{T}_i^a \epsilon(p_i)^\mu_b&=&-i f^{abc} \epsilon^\mu_c(p_i).\nonumber\\
\bold{T}_i^a \bar u_k(p_i)&=&T^a_{jk} \bar u_j(p_i).\nonumber\\
\bold{T}_i^a u_k(p_i)&=&-T^a_{kj}  u_j(p_i).
\eea

We are particularly interested in scattering amplitudes involving three final-state gluons ($ggg$) or the combination of a final-state quark-antiquark pair and a gluon ($q\bar qg$). 
With the definitions above, we can now evaluate the action of the operator given in eq.~\eqref{eq:GammaSoftDef} on such an amplitude.
\bea
&& {\bold \Gamma(\alpha_S(\mu^2),\{p_i \}, \epsilon)} \mathcal{A}_{ggg}\\
&&=\left[- C_A \Gamma_{\text{cusp.}}\left(\log\frac{-s_{12}}{\mu^2}+\log\frac{-s_{13}}{\mu^2}+\log\frac{-s_{23}}{\mu^2}\right)+\frac{3}{2} \gamma_c^A \right .\nonumber\\
&&\qquad \qquad \qquad \qquad \qquad \qquad \left.-\frac{C}{8} \left(\frac{\alpha_S(\mu^2)}{\pi}\right)^3  \left(C_A^3-24\frac{C_4^{AA}}{d_A T_A}\right)\right] \mathcal{A}_{ggg}.\nonumber\\
&& {\bold \Gamma(\alpha_S(\mu^2),\{p_i \}, \epsilon)} \mathcal{A}_{q\bar q g}\\
&&=\left[-\Gamma_{\text{cusp.}}\left(-(C_A-2C_F)\log\frac{-s_{12}}{\mu^2}+C_A \log\frac{-s_{13}}{\mu^2}+C_A\log\frac{-s_{23}}{\mu^2}\right) \right.\nonumber\\
&&\qquad \qquad \qquad \qquad \qquad \left.+\frac{1}{2} \gamma_c^A+ \gamma_c^F- \frac{C}{8} \left(\frac{\alpha_S(\mu^2)}{\pi}\right)^3  \left(C_A^3-24\frac{C_4^{AF}}{d_A T_F}\right) 
\right]\mathcal{A}_{q\bar q g}.\nonumber
\eea
Note that the formulae above are valid up to three loops.
The action of the soft anomalous dimension operator on our amplitudes is diagonal in color space such that the subtraction of infrared singularities becomes multiplicative.
We now want to make use of the factorization introduced in eq.~\eqref{eq:softfac} in order to simplify the subtraction of infrared poles.
By rewriting eq.~\eqref{eq:softfac} for finite amplitudes, we find
\bea
&&{\bold Z}(\alpha_S(\mu^2),\{p_i\},\epsilon) {\bold Z}_{\text{UV}} \mathcal{A}(\{p_i\},\epsilon)\nonumber\\
&&= \left[{\bold Z}_J(\alpha_S(\mu^2),\{p_i\},\epsilon) {\bold Z}_{\text{UV}}   {\bold J(p_4)} \right]\times\left[{\bold Z}(\alpha_S(\mu^2),\{p_i\},\epsilon) {\bold Z}_{\text{UV}}  \mathcal{F}_{1\to 23} \mathcal{A}_{1\to 23}^{(0)}\right].
\eea
The action of $\bold{\Gamma}$ on the amplitudes $\mathcal{A}_{1\to 23}$ is given by
\beq
{\bold \Gamma(\alpha_S(\mu^2),\{p_i \}, \epsilon)} \mathcal{A}_{1\to 23}=\left[- 2 C_R \Gamma_{\text{cusp.}}\log\frac{-s_{12}}{\mu^2} +\gamma_c^R \right] \mathcal{A}_{1\to 23}.
\eeq
Above, the sub- or superscript $R$ indicates the color representations of the colored particles of $\mathcal{A}_{1\to 23}$. 
This result can now be used in order to find 
\bea
&& {\bold \Gamma_J (\alpha_S(\mu^2),\{p_i \}, \epsilon)}  {\bold Z}_{\text{UV}}   {\bold J(p_4)} \\
&& =\left[C_A \Gamma_{\text{cusp.}}\left(\log\frac{-s_{12}}{\mu^2}-\log\frac{-s_{13}}{\mu^2}-\log\frac{-s_{23}}{\mu^2}\right)+\frac{1}{2} \gamma_c^A\right.\nonumber\\
&& \qquad \qquad \qquad \qquad \qquad \left.-\frac{C}{8} \left(\frac{\alpha_S(\mu^2)}{\pi}\right)^3  \left(C_A^3-24\frac{C_4^{AR}}{d_A T_R}\right)\right]  {\bold Z}_{\text{UV}}   {\bold J(p_4)} .\nonumber
\eea
Next, we perform the integration over $\mu^{\prime 2}$ in eq.~\eqref{eq:Zexp} and consequently obtain the necessary ingredients to remove all infrared singularities of the single-soft emission current. 
Indeed, we find that our results are finite once the subtraction procedure of eq.~\eqref{eq:subtdiv} is complete. 
The fact that the poles of the soft emission current computed from $\gamma^*\to q\bar q g$ amplitude and the $h\to ggg$ amplitude agree with the prediction based on the soft anomalous dimension matrix discussed here is a robust cross-check of our results.

%% file: Chapters/N4SYM.tex
\section{The soft current in $\mathcal{N}=4$ super Yang-Mills theory}
\label{sec:N4SYM}

Maximally super-symmetric Yang-Mills theory ($\mathcal{N}=4 $ sYM) is an excellent testing ground for many aspects of four-dimensional non-abelian gauge theory. 
It has countless times served as a laboratory to explore perturbation theory to perturbative orders simply not accessible in realistic theories like QCD.
One particular observation is that there is an interesting similarity between QCD and $\mathcal{N}=4$ sYM: 
The leading transcendental part of the perturbative expansion of certain quantities agrees among the two theories~\cite{Kotikov:2004er,Kotikov:2002ab}.
This correspondence even holds true for certain form factors of operators of the stress tensor multiplet~\cite{Jin:2018fak,Brandhuber:2017bkg,Brandhuber:2014ica,Brandhuber:2012vm}.
In particular, the form factor of three on-shell states $\Phi$ and the trace of two scalar fields $\phi$,
\beq
\mathcal{F}_2=\int d^d x \langle \Phi_1 \Phi_2\Phi_3 | \phi^I(x) \phi^I(x) |0\rangle,
\eeq
corresponds to the amplitude of a Higgs boson decaying to three gluons in QCD. 
This form factor has been of great interest to the community~\cite{Dixon:2022xqh,Dixon:2021tdw,Dixon:2020bbt,Dixon:2023kop,Yang:2019vag,Guan:2023gsz,Lin:2021qol,Lin:2021kht} and was recently computed to staggering eight-loop accuracy in the planar limit of the theory~\cite{Dixon:2022rse}.

Similar to the QCD case discussed above, the soft limit of these form factors can be used to extract the soft current in $\mathcal{N}=4$ sYM theory. 
To achieve this, we start by using the integrand for the form factor determined in ref.~\cite{Lin:2021kht} at two- and three-loop order.
We then apply our integrand expansion technology and compute the first term in the maximally soft limit of this form factor. 
We obtain a pure function (i.e. of maximal transcendental weight) for both the two- and three-loop result.
We then compare our result with the maximally soft limit of the decay form factor of a Higgs to three gluons in QCD. 
Indeed, we find that these two results agree to all orders in the dimensional regulator for the leading transcendental contribution.
Consequently, we determine that the single-soft emission current in $\mathcal{N}=4$ sYM theory is identical to the leading transcendental part of the QCD results quoted above. 
This validates the principle of maximal transcendentality for the quantity computed here.

We find that the maximally soft limit of the $\mathcal{F}_2$ form factor at three-loop order relative to its Born contribution can be cast in the following form.
\beq
\mathcal{F}^{(3)}_2/
\mathcal{F}^{(0)}_2=\frac{a_S^3 \pi^3 i}{ 2^{11}\times 3^6\times 5\times \epsilon^6} \left(\frac{(-2p_2p_4)(-2p_3p_4)}{(-2p_2p_3) \mu^2}\right)^{-3\epsilon} \left[C_A^3 F^{(3),\text{P}}+\frac{C_4^{AA}}{d_A T_A}F^{(3),\text{NP}}\right].
\eeq
In the equation above, we defined two uniform transcendental functions that we present here as an integer-linear combination of our canonical soft master integrals defined in eq.~\eqref{eq:CanMsDef}. 
We would like to emphasize that the solution of the canonical soft master integrals we provide is valid only up to transcendental weight 8.
In contrast, the expressions below are correct to arbitrary order of the Laurent expansion in $\epsilon$.
\bea
F^{(3),\text{P}}&=&
-3996 M_c^{50}+4032 M_c^{49}+1350 M_c^{48}-8640 M_c^{47}-12960 M_c^{46}+3807 M_c^{45}\\
&&+6156 M_c^{43}+720 M_c^{41}+702 M_c^{40}+5400 M_c^{39}-3240 M_c^{37}-1125 M_c^{36}-6570 M_c^{35}\nonumber\\
&&+360 M_c^{33}-12960 M_c^{31}-540 M_c^{28}-2376 M_c^{26}+1890 M_c^{25}+5184 M_c^{24}\nonumber\\
&&+9720 M_c^{23}+16200 M_c^{21}-560 M_c^{20}+8076 M_c^{19}-120 M_c^{18}-116640 M_c^{17}\nonumber\\
&&-1944 M_c^{16}-4860 M_c^{15}-180 M_c^{13}+103680 M_c^{12}-936 M_c^{11}+16200 M_c^{10}\nonumber\\
&&-19440 M_c^9-378 M_c^8-7344 M_c^7-3240 M_c^6+6480 M_c^5+864 M_c^4-15552 M_c^1.\nonumber
\eea
\bea
F^{(3),\text{NP}}&=&
95904 M_c^{50}-96768 M_c^{49}-32400 M_c^{48}-103680 M_c^{47}-91368 M_c^{45}-147744 M_c^{43}\nonumber\\
&&-16848 M_c^{40}-129600 M_c^{39}+77760 M_c^{37}+27000 M_c^{36}+157680 M_c^{35}+4320 M_c^{33}\nonumber\\
&&+12960 M_c^{28}+57024 M_c^{26}-45360 M_c^{25}+62208 M_c^{24}-233280 M_c^{23}+67392 M_c^{21}\nonumber\\
&&+13440 M_c^{20}+75744 M_c^{19}+2880 M_c^{18}+46656 M_c^{16}+4320 M_c^{13}+22464 M_c^{11}\nonumber\\
&&+77760 M_c^{10}+466560 M_c^9+9072 M_c^8+176256 M_c^7+77760 M_c^6-155520 M_c^5\nonumber\\
&&+10368 M_c^4.
\eea
Above we introduced
\beq
d_A=(n_c^2-1),\hspace{1cm} T_A=n_c.
\eeq

%% file: Chapters/SoftXS.tex
\section{Single-real threshold contribution to the Higgs boson and Drell-Yan production cross sections at N${}^4$LO}
\label{sec:thresholdxs}

The inclusive gluon fusion Higgs boson production cross section and the Drell-Yan production cross section of an electron-positron pair are some of the most important LHC observables. Currently, their predictions are known through N$^3$LO in perturbative QCD~\cite{Duhr:2021vwj,Duhr:2020seh,Mistlberger:2018etf,Anastasiou:2015ema}.
Going beyond the current state of the art is a formidable challenge and we present here a first contribution towards this step.

The LHC cross sections for the production of a virtual photon or a Higgs boson in gluon fusion in the infinite top quark mass limit is described by the following factorization formula.
\beq
\label{eq:hadronicxs}
\sigma_B=\tau \hat \sigma_0^BC^2_B \sum_{ij} f_i(\tau) \circ_\tau \eta_{ij}^B(\tau) \circ_\tau f_j(\tau) ,\hspace{1cm} B\in\{H,\gamma^*\}.
\eeq
Above, the $f_i$ are parton distribution functions (PDFs), $\hat \sigma_0^B$ represents the partonic Born cross section, and we define the ratio $\tau=Q^2/ S$, such that $Q$ is the virtuality of the virtual photon or the mass of the Higgs boson, and $S$ is the hadronic center-of-mass energy. 
The PDFs are convoluted with the partonic coefficient functions using standard Mellin convolutions indicated by the symbol $\circ$.
The partonic coefficient functions $\eta^B_{ij}$ are given by
\beq
\label{eq:PCF}
\eta^B_{ij}(z)=\frac{\mathcal{N}_{ij}}{2Q^2 \hat \sigma_0^B}  \sum_{m=0 }^\infty \int d \Phi_{B+m}\mathcal{M}_{ij\to B+m}.
\eeq
The  normalization factor $\mathcal{N}_{ij}$ depends on the initial state and is given by
\beq
\mathcal{N}_{gg}=\frac{1}{4(n_c^2-1)^2(1-\epsilon)^2},\hspace{1cm}
\mathcal{N}_{q\bar q}=\frac{1}{4n_c^2},
\eeq
where $g$, $q$ and $\bar q$ represent a gluon, quark and anti-quark respectively, and $n_c$ denotes the number of colors. The coefficient $C_B$ is simply unity for the production cross section of a virtual photon and equal to the Wilson coefficient~\cite{Chetyrkin:1997un,Schroder:2005hy,Chetyrkin:2005ia,Kramer:1996iq} for the effective field theory describing the interactions of a Higgs boson with gluons in the limit of infinitely large top quark mass~\cite{Inami1983,Shifman1978,Spiridonov:1988md,Wilczek1977}. 
The color and spin summed squared matrix element is given by $\mathcal{M}_{ij\to B+m}$.
This squared matrix element describes the production of the desired boson $B$ and $m$ final state partons in the collision of initial state partons $i$ and $j$. 
In this article, we focus in particular on the contribution for one final state gluon (i.e. $m=1$).
We refer to this contribution as the single real emission (R) contribution to the inclusive cross section.
The corresponding partonic coefficient function is consequently given by
\beq
\label{eq:RPCF}
\eta^{B,R}_{ij}(z)=\frac{\mathcal{N}_{ij}}{2Q^2 \hat \sigma_0^B}  \int d \Phi_{B+1}\mathcal{M}_{ij\to B+1}.
\eeq
We focus on the limit in which the energy of the final state parton vanishes. 
This limit is referred to as the production threshold, as all energy of the colliding partons is used to produce the final state boson. 
To parametrize this limit, we introduce the following variables
\beq
\bar z=1-z,\hspace{1cm}z=\frac{Q^2}{s}.
\eeq
The threshold (or soft) limit is given by $\bar z\to 0$.
We can now exploit the factorization scattering amplitudes as introduced in eq.~\eqref{eq:softfac} to compute the threshold limit of the single-real emission partonic coefficient function.
\beq
\eta^{B,R,\text{thr.}}_{ij}(z)=\lim\limits_{\bar z \to 0} \eta^{B,R}_{ij}(z)=\frac{\mathcal{N}_{ij}}{2Q^2 \hat \sigma_0^B}  \int d \Phi_{B+1} \sum\limits_{\text{Spin, Color}} \Big| \bold{J}(p_g) \mathcal{A}_{p_ip_j\to B}\Big|^2.
\eeq
The result for this part of the partonic coefficient function can be expanded in the strong coupling constant (eq.~\eqref{eq:asdef}).
\beq
\eta^{B,R,\text{thr.}}_{ij}(z)= \sum_{o=0}^\infty a_S^o \eta^{B,R,\text{thr.}, (o)}_{ij}(z).
\eeq
The above single-real emission contribution to the partonic coefficient function computed through N$^4$LO in perturbative QCD represents a major result of this article.
To obtain this result through three loops in QCD, we make use of our newly derived results for the soft current (eq.~\eqref{eq:currentstruc}) and apply it to the purely virtual amplitudes of the scattering process in question. 
These virtual amplitudes are currently available in the literature to four-loop order~\cite{Lee:2022nhh,Agarwal:2021zft,Lee:2021uqq,vonManteuffel:2020vjv,Gehrmann:2010ue,Gehrmann:2010tu,Baikov:2009bg,Gehrmann2005}, even beyond what is required here. 
To arrive at the desired result for the partonic coefficient function through N$^4$LO, we first perform an analytic continuation of the soft current in eq.~\eqref{eq:coefdef} into the production region. 
Next, we apply the current to the purely virtual amplitudes to obtain the threshold limit of the desired scattering amplitude. 
Then, we interfere the soft scattering amplitude with its complex conjugate and finally perform the integration over the single emission phase space $d \Phi_{B+1}$. 

We express our results in terms of Laurent expansions in the dimensional regulator $\epsilon$ and express threshold singularities in terms of standard Dirac delta functions and plus distributions. 
Their action on a test function $f(\bar z)$ is given by
\beq
f(0)=\int_0^1 d\bar z \delta(\bar z) f(\bar z), \hspace{1cm} \int_0^1d\bar z \left[\frac{\log^n \bar z}{\bar z}\right]_+ f(\bar z)=\int_0^1 d\bar z \frac{\log^n \bar z}{\bar z} (f(\bar z)-f(0)).
\eeq
In order for our results to be usable for the computation for the N$^4$LO production cross section we truncate the Laurent expansion in $\epsilon$ 
at  $\mathcal{O}(\epsilon^{8-2n})$ at N$^n$LO in QCD perturbation theory for $n\in\{1,2,3,4\}$.
Note, that first approximate results for the full threshold limit of the N$^4$LO production cross section already appeared in refs.~\cite{Duhr:2022cob,Das:2020adl}.
We confirm previous computations through N$^3$LO in the literature~\cite{Anastasiou:2012kq,Dulat:2014mda,Anastasiou:2013mca,Kilgore:2013gba,Harlander:2002wh,Anastasiou2002,Ravindran:2003um}.
We present our results in terms of computer readable files in association with the arXiv submission of this article.

%% file: Chapters/Conclusion.tex
\section{Conclusion}
\label{sec:conclusions}
In this article, we computed N$^3$LO corrections to the single-soft emission current applied to amplitudes with two color-charged partons. 
Our result is a significant contribution to our understanding of soft factorization at N$^3$LO in perturbative QFT and the understanding of infrared singularities at N$^4$LO.

We achieved our results by performing a systematic expansion of three-loop scattering amplitudes involving one color-neutral boson and three partons around the limit of one gluon becoming soft. To facilitate this expansion, we develop a new method for the systematic expansion of Feynman graphs around soft limits. 
We emphasize the generality of this technique and apply it to obtain our results for the soft emission current as a first example.

We perform the expansion of scattering matrix elements in QCD and in maximally super-symmetric Yang-Mills theory. 
We observe that the result from the two different gauge theories agree at the highest transcendentality, in accord with previous conjectures. Furthermore, we use our new results to determine the contributions to the threshold approximation of the Higgs boson and Drell-Yan production cross sections at the LHC at N$^4$LO in perturbative QCD. To facilitate the use of our results, we make them available in terms of computer readable files associated with the arXiv submission of this article.

{\bf{Note:} }During the completion of this manuscript a separate calculation of the three-loop single-soft emission current became public on the arXiv in ref.~\cite{Chen:2023hmk}. 
We have found complete agreement among our main results for coefficients $K_2^{(3)}$ (eq.~\eqref{eq:K23}), $K_{4A}^{(3)}$ (eq.~\eqref{eq:K4A3}) and $K_{4F}^{(3)}$ (eq.~\eqref{eq:K4F3}). 
It is worth noting that the methods employed in ref.~\cite{Chen:2023hmk} and in this article are substantially different, and obtaining matching results thus provides a robust cross-validation.